\documentclass[11pt,a4paper]{article}

\usepackage{microtype}
\usepackage[utf8]{inputenc}
\usepackage{jcappub}
\usepackage{color}
\usepackage{afterpage}
\graphicspath{{img/}}

\usepackage[]{hyperref}
\hypersetup{final}
\makeatletter
\AtBeginDocument{%
    \hypersetup{%
      pdfinfo={%
        Title={\@title},
        Subject={},
        Author={},
        Keywords={\@keywords},
        },
      pdfnumcopies=2
    }
}
\makeatother

\usepackage[obeyFinal]{todonotes}
\usepackage[markifdraft,dirty=*]{gitinfo2}

\toccontinuoustrue

\newcommand{\psiend}{\ensuremath \psi_{\text{end}}}
\newcommand{\psin}{\ensuremath \psi_{N}}
\newcommand{\mpl}{\ensuremath M_\text{Pl}}

\newcommand{\phin}{\ensuremath \phi_{N}}

\title{Expectations for Inflationary Observables: Simple or Natural?}
\author{Nathan Musoke,}
\emailAdd{n.musoke@auckland.ac.nz}
\author{Richard Easther}
\emailAdd{r.easther@auckland.ac.nz}
\affiliation{Department of Physics, University of Auckland, Private Bag 92019, Auckland, New Zealand}

\abstract{%
We describe the general inflationary dynamics that can arise with a single, canonically coupled field where the inflaton potential is a 4-th order polynomial. This scenario yields a wide range of combinations of the empirical spectral observables, $n_s$, $r$ and $\alpha_s$. However, not all combinations are possible and  next-generation cosmological experiments have the ability to rule out all inflationary scenarios based on this potential. Further, we construct inflationary priors for this potential based on physically motivated choices for its free parameters. These can be used to determine the degree of tuning associated with different combinations of $n_s$, $r$ and $\alpha_s$ and will facilitate treatments of the inflationary model selection problem. Finally, we comment on the implications of these results for the naturalness of the overall inflationary paradigm. We argue that ruling out all simple potentials would not necessarily imply that the inflationary paradigm itself was unnatural, but that this eventuality would increase the importance of building inflationary scenarios in the context of broader paradigms of ultra-high energy physics.}

\keywords{Inflation, cosmology, observables, microwave background, naturalness}

\arxivnumber{1709.01192}

\begin{document}

\maketitle

\section{Introduction}
\label{sec:introduction}

Inflation has been part of  the conventional cosmological narrative of the evolving universe since the early 1980s \cite{Guth:1980zm,Linde:1981mu,Albrecht:1982wi} and can  account for key observations, including the  overall flatness, isotropy, homogeneity of the universe, and the nearly scale-invariant spectrum of perturbations. However, there is no clear and compelling ``standard model'' of inflation, making it difficult to formulate conclusive tests of the overall paradigm.

There are two broad approaches to establishing  generic  properties of the  inflationary phase within the  framework of slow-roll inflation driven by a single scalar field with a canonical kinetic term.
The first stipulates that the inflationary potential is algebraically {\em simple}, with monotonic derivatives and an elementary algebraic form \cite{Boyle:2005ug,Bird:2008cp,Boyle:2008ri,Ijjas:2013vea}; the quadratic and quartic potentials are the canonical examples.
For simple scenarios the tensor to scalar ratio $r$ is typically $r \gtrsim 0.1$, while values of $r\lesssim 0.01$ appear fine-tuned.
However, relatively large values of $r$ suggest that the inflaton field makes a super-Planckian excursion during the course of inflation \cite{Lyth:1996im}.
At large VEVs generic Planck scale operators  contribute significantly to the potential; these contributions must be fine-tuned if the potential is smooth and flat  over a trans-Planckian field range, a violation of technical {\em naturalness\/}.
From this perspective, a very small ($r\ll10^{-3}$) tensor background is the more likely outcome. As a consequence,  {\em simple\/} and {\em natural\/} are far from  synonymous in the context of  inflation. The tension between these viewpoints is resolved if a symmetry suppresses  Planck-scale and  other very high energy corrections to the  potential.
Within string theory, this approach leads to the identification of scenarios such as monodromy inflation  \cite{Baumann:2014nda,Silverstein:2008sg,McAllister:2008hb,Flauger:2009ab} which generate a significant tensor spectrum in a framework whose stability against  ultraviolet corrections  can be directly assessed.
Likewise, an effective single-field model can arise as a cooperative effect among  many individual fields, each of which makes a sub-Planckian excursion  \cite{Liddle:1998jc,Copeland:1999cs,Kanti:1999vt,Kanti:1999ie,Dimopoulos:2005ac,Easther:2005zr}.

The status of inflationary models is usefully discussed from a Bayesian standpoint \cite{Easther:2011yq,Planck:2013jfk,Martin:2013nzq,Ade:2015lrj}.
Rather than performing parameter estimation for  empirical observables associated  with the primordial perturbations such as $n_s$ (spectral index), $r$, $A_s$ (amplitude), $\alpha_s$ (running) or $f_{NL}$ (non-Gaussianity),  candidate inflationary models can be defined by the {\em prior}  specifying  the functional form of the model and the distributions from which its free parameters are drawn.
In many cases a natural measure on the inflationary parameter space  map to highly non-uniform and strongly correlated distributions for  $n_s$, $r$ and $\alpha_s$  \cite{Easther:2011yq,Adshead:2010mc}.
Furthermore, Bayesian methodology addresses the  inflationary model selection problem via the  {\em evidence\/} for each identified inflationary scenario.
For a  model with an $m$-dimensional parameter vector $\bar a$ drawn from a joint distribution $P(\bar a)$ the evidence is  
 \begin{equation} \label{eq:evidence}
 E= \int da^m P(\bar a) {\cal L}(\bar a) 
 \end{equation}
 where ${\cal L}(\bar a) $ is the likelihood derived from a specified set of observations and $P(\bar a)$ is normalised to unity \cite{Jenkins:2011va,Easther:2011yq,Planck:2013jfk,Martin:2013nzq,Ade:2015lrj}.  If two models $M_1$ and $M_2$ are {\em a priori} equally likely,  $E_1/E_2$ is the odds ratio for the two scenarios; if the relative prior likelihood of $M_1$ and $M_2$ is $P(M_1) / P(M_2)$ then the odds ratio becomes $E_1 P(M_1)/E_2 P(M_2)$.
 
Quantitative treatments of inflationary model selection typically assume that  the $P(M_j)$ are equal, where $j$ labels  the models under consideration.
The statement that a model is fine-tuned can reflect external knowledge regarding the model's physical origin, resulting in a stipulation that the corresponding $P(M_i)$ is small.
Alternatively, fine-tuning may become apparent when parameter estimation reveals that the posteriors of one or more free parameters differ greatly from a physically motivated prior.
In some cases this tuning, and thus the impact on $P(M_j)$, can be quantified.
For example, given a ``new physics scale'' $\Lambda$ it is well-known that the  effective mass $m^2$ of a scalar field with bare mass  $m^2_b$ is
 \begin{equation}
 m^2 = m^2_b + f_2 \Lambda^2 
 \end{equation}
where $f_2$ is expected to be of order unity  \cite{Baumann:2014nda}.
With $\Lambda \sim \mpl$, a uniform distribution for $f_2$, and in the absence of an underlying symmetry controlling $f_2$ it can appear that $P(M_{\tiny \rm quad }) \sim 10^{-11}$ since successful quadratic inflation  requires $m^2 \sim 10^{-11} \mpl^2$.

While attempts to identify truly canonical attributes of the inflationary paradigm have been largely futile, the accuracy and precision of measurements of the primordial perturbations has improved dramatically over the last 25 years \cite{Adam:2015rua,Ade:2015xua,Ade:2015lrj}.
In particular, current constraints on $n_s$ and $r$ barely with overlap the region of parameter space open to the quadratic model, while the quartic potential has been disfavoured by observations since the first WMAP data release \cite{Peiris:2003ff}.
It is thus unlikely that very simple implementations of the inflationary paradigm are consistent with observational constraints.
In Bayesian terms, this allows us to update inflationary priors in light of data.
Consequently, although we cannot draw inferences about the detailed structure of high energy physics, we can state with increasing confidence that (for example) it is unlikely our universe underwent a period of inflation driven by a potential that is quadratic or steeper.

The immediate goal of this paper is to fully assess the possible observable consequences of inflation due to a single scalar field with a 4-th order polynomial.
This is the most general renormalizable potential for a minimally coupled scalar field.
Moreover, this  is the simplest polynomial that supports inflection point inflation \cite{Okada:2016ssd,Allahverdi:2006iq,Lyth:2006ec,BuenoSanchez:2006rze,Martin:2013nzq} while being bounded below, so it yields both large field and small field inflation \cite{Dodelson:1997hr}.
Clearly this is not a new model; to our knowledge this scenario was first explicitly analysed in 1990 \cite{Hodges:1989dw,Hodges:1990bf}, with a focus on its ability to support designer scenarios with broken scale invariance in the large-field regime.
In addition,  Ref.~\cite{Bird:2008cp} gives a Monte Carlo  sampling of the parameter space, Ref.~\cite{Destri:2007pv} presents an analysis of possible observable outcomes with constraints from early WMAP data, Ref.~\cite{Aslanyan:2015hmi} provides up-to-date constraints on potential parameters, and Ref.~\cite{Fowlie:2016jlx} performs a Bayesian analysis in the context of the hierarchy problem.

Our treatment adds to previous discussions by a) presenting a more complete analysis of the possible observables including the spectral running; b) stating clear ``inflationary priors'' that will facilitate parameter estimation and model selection calculations; c) treating these model specifications as hyperpriors for a generative model \cite{Price:2015qqb} of the spectrum and computing distributions for the usual spectral parameters, showing that these are highly nonuniform; d) providing a qualitative analysis of the extent to which these inflationary priors can be updated with reference to present-day data, and e) showing that data from plausible future experiments could potentially rule out {\em all} inflationary scenarios based on the single-field 4-th order potential.

In particular, although the catalog of  inflationary behaviours available to a 4-th order polynomial potential  populates a substantial subset of the $\{n_s,r,\alpha_s\}$ parameter-space, a significant fraction of this region is revealed to be incompatible with the potential and near-future experiments will have the ability to exclude the full  parameter space.
This is a  key result of this paper:  the 4-th order polynomial -- the most complex potential compatible with the tree-level action of a  single, renormalizable, minimally coupled scalar field -- can  be falsified by experiments that are part of the current roadmap for experimental cosmology. 
The physical basis of this result is that parameter combinations that predict smaller values of $r$ tend to have larger values of $\alpha_s$ \cite{Boyle:2008ri,Adshead:2010mc,Munoz:2016owz}. We do not find scenarios where both parameters are vanishingly small and $n_s$ lies inside its currently permitted range, so even null results for both $r$ and $\alpha_s$ could rule out a 4-th order potential.

This analysis sets the stage for a broader discussion of the status of fine-tuning in inflationary cosmology.
Our results allow us to construct priors based on a variety of physically motivated expectations for values of parameters in the potential, from which we derive distributions for $n_s$, $r$ and $\alpha_s$.
These are typically highly non-uniform and effectively quantify the degree of tuning associated with different values of the empirical spectral parameters.
Furthermore, in some regions of parameter space the pre-inflationary initial conditions must themselves be chosen carefully in order for inflation to begin, further heightening the degree of tuning these models require.

The prospect of all putatively simple models  being ruled out by observations  forces a closer evaluation of the naturalness of  these scenarios. Given that even simple potentials are more plausible when embedded in more complex theories we argue that this development need not immediately undermine confidence in the overall inflationary paradigm. However, if the simplest potentials are   ruled out, arguments that inflation is natural must increasingly be framed within the context of more general discussions about the properties and nature of ultra-high energy physics.

\section{Inflationary Dynamics}
\label{sec:the_model}

 A minimally coupled scalar field with a canonical kinetic term and potential $V(\phi)$ obeys the Klein-Gordon equation, 
\begin{equation}
\ddot{\phi}+3H\dot{\phi}+V'(\phi) = 0\,,
\end{equation}
where $H$ is the Hubble parameter. The scale factor $a(t)$ is described by the usual Friedman equation
\begin{equation}
H^2 = \left(\frac{\dot{a}}{a}\right)^2 = \frac{1}{3\mpl^2}\left[\frac{1}{2}\dot{\phi}^2 + V(\phi)\right]\, ,
\end{equation}
where $\mpl$ is the (reduced) Planck mass.
Our analysis is based on the generic quartic potential, the most complicated renomalizable tree-level action a single scalar field can possess
\begin{equation}
V(\phi) = \sum_{i=0}^4 c_i \phi^i\, .
\end{equation}
We require that the potential has a global, stable minimum at $\phi_\text{min}$ with $V(\phi_\text{min}) = 0$; a linear shift can always give $\phi_\text{min} = 0$.
As a consequence of these criteria we fix $c_0 = c_1 = 0$, $c_2 > 0$ and $c_4 >0$ so
\begin{equation}
    \label{eq:v_phi}
	V(\phi)
	=
	\frac{m^2}{2} \phi^2 - \frac{g}{3} \phi^3 + \frac{\lambda}{4} \phi^4
\end{equation}
with $\lambda > 0$.
We  also choose $g \ge 0$ without loss of generality since the underlying theory is invariant under $\phi \rightarrow -\phi$ if $g \rightarrow -g$.

Inflationary observables may be  characterised by the potential slow roll parameters \cite{Liddle:1994dx}
\begin{align}
    \label{eq:epsilon_approx}
    \epsilon
    &=
    \frac{\mpl^{2}}{2} \left( \frac{V'}{V} \right)^2 \, ,
    \\
    \label{eq:eta}
    \eta
    &=
    \mpl^{2} \frac{V''}{V} \, ,
    \\
    \label{eq:xi}
    \xi
    &=
    \mpl^{4} \frac{V'}{V} \frac{V'''}{V} \,    .
\end{align}
Inflation occurs when $\epsilon \lesssim 1$.
The slow roll parameters specify the inflationary observables,
\begin{align}
    \label{eq:n_s}
    n_s - 1 
        &\approx -6 \epsilon + 2 \eta  \, ,
    \\
    r
    &\approx
    16 \epsilon  \, ,
    \\
    \alpha_s
    &\approx
    16 \epsilon \eta - 24 \epsilon^2 - 2 \xi \, .
\end{align}
These quantities are evaluated at the field value $\phin$ corresponding to the instant at which the corresponding mode leaves the horizon, $N$ $e$-folds before the end of inflation, with
\begin{equation}
    \label{eq:N_integral}
    N
    =
    \frac{1}{\mpl^2}
    \int_{\phi_{end}}^{\phin}
    \frac{V(\phi)}{V'(\phi)}
    d \phi
    .
\end{equation} 

A 4th-order potential can have up to three local extrema.
Given that $\lambda > 0$, $V(\phi)$ may have either two local minima (including a global minimum) and a local maximum, a global minimum and a saddle point, or a single, global minimum.
The ratio $M_{Pl} V'/V$ goes to zero at large $\phi$ so this potential certainly supports inflation at very large field values.
However, the  potential can  have an inflection point at arbitrary values of $\phi$, in the vicinity of which $V'/V$  is small enough to support inflation.
Consequently, there are two distinct inflationary regimes -- the large-field case and a possible small field case around an inflection point -- and we analyse them separately.

\section{Small Field: Observables From Inflection Point Inflation}
\label{sec:analysis}

Inflection point potentials have been studied in many contexts.
Some are motivated by particle physics, e.g. an inflection point added to monomial chaotic models through radiative corrections \cite{Ballesteros:2015noa},  Higgs potentials  \cite{Okada:2016ssd}, the minimal supersymmetric standard model (MSSM) \cite{Allahverdi:2006iq,Lyth:2006ec,BuenoSanchez:2006rze,Martin:2013nzq}, D-brane inflation \cite{Baumann:2007np,Baumann:2007ah}, or loop inflection-point inflation \cite{Dimopoulos:2017xox}.
The 4-th order model looked at here is discussed by Martin, Ringeval and Vennin \cite{Martin:2013tda}, who dub it Renormalizable Inflection Point Inflation.\footnote{Section 4.18 of Ref~ \cite{Martin:2013tda} treats the $\delta=0$ case (in our notation) while Section~5.7 discusses the general $\delta\ne0$ case; we give a unified account of both situations and our conclusions regarding the tuning required for inflation and the overall falsifiability of the scenario are new.}

We begin by noting that the potential \eqref{eq:v_phi} has an exact saddle point at  $\phi = g/2\lambda$ when $m^2 = \frac{g^2}{4 \lambda}$ and reparameterize  via
\begin{equation}
	m^2 = \frac{g^2}{4 \lambda} (1 + \delta) \, .
	\label{eqn:mmpert}
\end{equation}
If the dimensionless parameter $\delta$ is negative we have a trapping potential\footnote{Such a potential would support hilltop inflation but this possibility does not change any of our conclusions as it yields an $n_s$ even more inconsistent with observations -- see for example \cite[Section~5.7, Figure~125]{Martin:2013tda}.} and $\delta =0$ is the saddle point limit. Our first task is to determine the overall inflationary region inside the potential. The slope $V'$ is minimized at
\begin{equation}
	\phi_\text{inf}
	=
	\frac{g \bigl( 2 \pm \sqrt{1-3\delta} \bigr)}{6 \lambda} \, .
	\label{}
\end{equation}
The positive root coincides with the  saddle  location when $\delta=0$ and this inflection point only exists if $\delta<1/3$.
We will see that a realistic inflationary phase requires $\delta \ll 1/3$, so we  write
\begin{equation}
	\phi_\text{inf}
	\approx
	\frac{g}{2 \lambda} \Bigl(1 - \frac{\delta}{2}\Bigr)
    =
	M \Bigl(1 - \frac{\delta}{2}\Bigr)
	\label{}
\end{equation}
and make the identification $g = 2 \lambda M$.
Changing variables to $\psi = \phi - \phi_\text{inf}$ and dropping terms beyond first order in $\delta$ gives
\begin{equation}
	V(\psi)
	=
    \frac{\lambda}{12} \left[
        M^{4} (1 + 6 \delta) + 
        12 M^{3} \delta \psi + 
        2 M \psi^{3} \left(- 3 \delta + 2\right) + 
        3 \psi^{4}
    \right]
	.
	\label{potential_psi}
\end{equation}
Note that the quadratic term would be proportional to $\delta^2$ and is thus absent in this approximation. The overall minimum lies at exactly $\psi=-M$ with $V(-M)=0$.
The shape of the potential is determined by the  parameters $M$ and $\delta$, while its overall scale is fixed by $\lambda$.
Since inflation occurs when $\phi \approx M$ the total field excursion is of order $M$.
In this section we assume $M \le \mpl$, i.e. small field inflation.
The $\psi^4$ term  in equation~(\ref{potential_psi})  ensures  the potential is bounded below. 

During a viable inflationary phase $V'/V$ and $V''/V$ are both small. Dropping higher order terms in $\delta$ and noting that $V$ changes much more slowly than $V'$ and $V''$ near $\psi=0$, we  approximate the slow roll parameters as
\begin{align}
    \label{eq:epsilon_approx_inf}
    \epsilon
    &        \approx
    \frac{72 \mpl^{2}}{M^2} \left( \delta + \frac{\psi^2}{M^2} \left( 1 - \frac{3}{2} \delta \right)\right)^{2}
    \\
    \label{eq:eta_inf}
    \eta
    &     \approx
    12 \frac{\mpl^{2} \psi }{M^{3}} \left(2 - 3 \delta\right)
    \\
    \label{eq:xi_inf}
    \xi
    &
    \approx
    288 \delta \frac{\mpl^{4}}{M^{4}} \left(1 + 3 \frac{\psi}{M}\right)
    \,.
\end{align}
To have any inflation at all we need $\epsilon<1$  at $\psi=0$, so
\begin{equation}
    \label{eq:delta_max}
    \delta 
    \lesssim 
    \frac{\sqrt{2}}{12} \frac{M}{\mpl} 
    \equiv \delta_{\text{max}}\, .
\end{equation}
If $M \lesssim \mpl$ this immediately implies $\delta \lesssim 0.1$, justifying our decision to retain only terms first order in $\delta$. 

\begin{figure}[tpb]
    \centering
    \includegraphics[width=0.8\linewidth, draft=false]{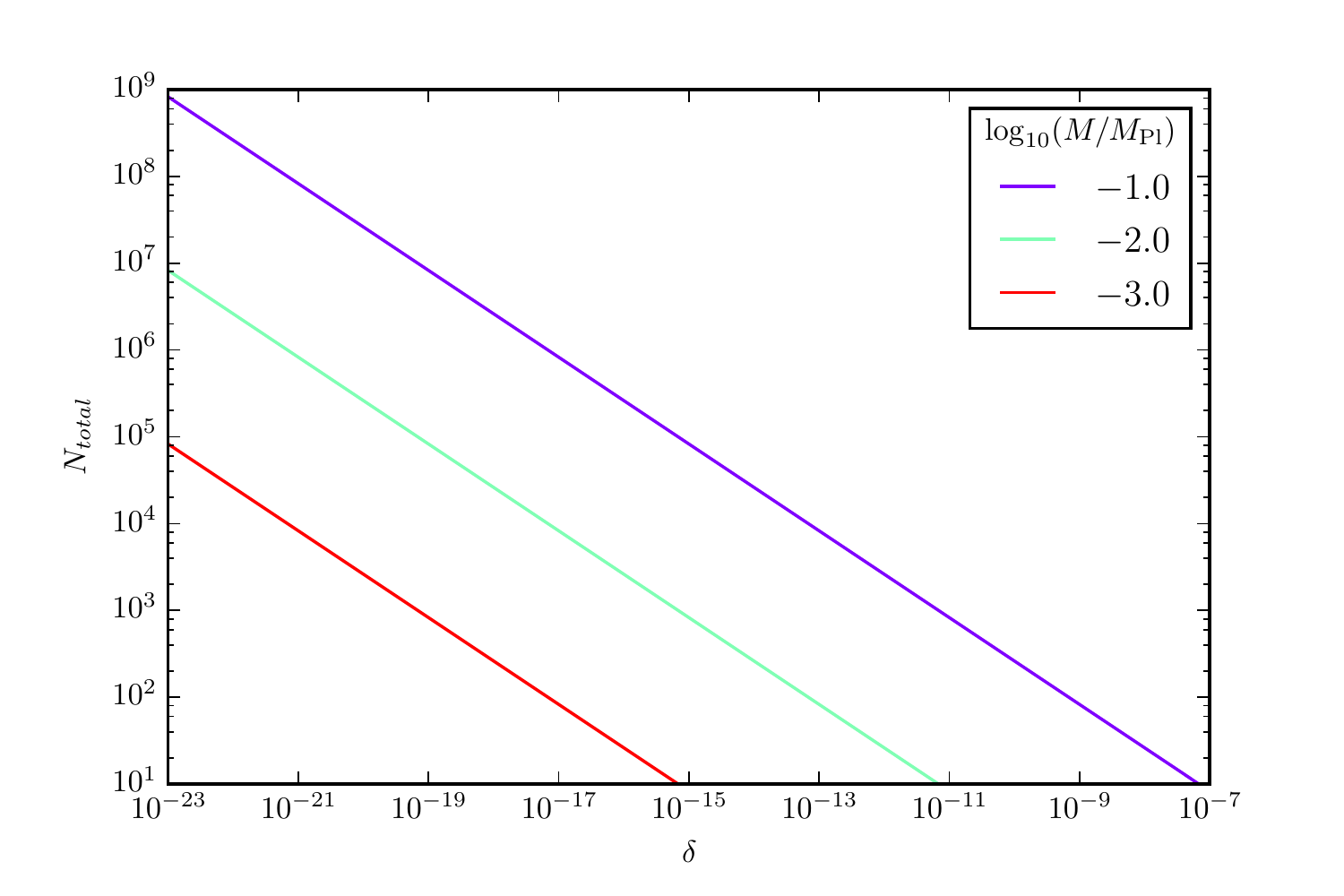}
    \caption{%
        The total number of $e$-folds of inflation $N$ as function of $\delta$ for representative values of $M$, found by numerically integrating~\eqref{eq:N_integral}.
        The value of $\delta$ required for a fixed number of $e$-foldings scales as $(M/\mpl)^4$, consistent with the constraint~\eqref{eq:delta_max2}.
    }
    \label{fig:N_total}
\end{figure}

We also need the higher order slow roll parameters to be small.  By definition, $V''=0$ at an inflection point, so $\eta=0$ when $\psi=0$ and we do not find a new constraint. However, writing $\delta = c\delta_{\text{max}}$ and substituting into $\xi$ we find
\begin{equation}
\xi \approx +c \frac{24 \sqrt{2} \mpl^3}{M^3} \, .
\end{equation}
The full hierarchy of slow roll equations (in either the potential or Hubble slow roll formalism \cite{Liddle:1994dx}) shows that $d \epsilon /d N$ and $d \eta /d N$ are large if $\xi \ge 1$ so the inflationary region of the potential is very narrow and the number of $e$-folds $N$  less than unity. Consequently, $N$ and $\alpha_s$ are  correlated in generic slow roll models \cite{Easther:2006tv}; current observational bounds on $\alpha_s$ and ensuring  $N \gtrsim 30$ both require $|\xi| \lesssim 0.01$ at $\psi=0$, or 
\begin{equation}
c \lesssim  \frac{0.01}{24 \sqrt{2}} \left(\frac{M}{\mpl}\right)^3 \, .
\end{equation}
Since $\delta = c  \delta_{\text{max}}$, combining with equation~\eqref{eq:delta_max} yields
\begin{equation}  \label{eq:delta_max2}
\delta \lesssim \frac{0.01}{288} \left(\frac{M}{\mpl}\right)^4 \,;
\end{equation}
decreasing $M$ by an order of magnitude decreases this upper bound on $\delta$ by four orders of magnitude.
If $M \sim 10^{-2} \mpl$ the cubic term is tuned to parts in $10^{12}$, independently of the constraint on $\lambda$ required to produce a suitable amplitude for the perturbation spectrum.
This relationship is illustrated in Figure~\ref{fig:N_total}.
Similar tuning requirements have been seen in the analysis of other models with inflection points, including MSSM motivated potentials \cite{BuenoSanchez:2006rze} and accidental inflation \cite{Linde:2007jn}.
This fine tuning can be reduced by adding a constant term to the potential~\eqref{eq:v_phi}, decreasing $V'/V$ and effectively flattening the potential \cite{Enqvist:2010vd,Hotchkiss:2011am}.
The trade-off is that $V(\phi_{\text{min}}) > 0$, making the model unrealistic in the absence of a mechanism to subsequently ensure that the vacuum energy becomes vanishingly small.

To compute $N$ note that inflation begins and ends when $\epsilon \approx 1$, or 
 \begin{align}
     \label{eq:psi_end_sq}
     \psiend^{2}
     &\approx
     \frac{M^2}{1 - \frac{3}{2}\delta}
     \left(
         -\delta \pm \frac{\sqrt{2}}{12} \frac{M}{\mpl} (1+6\delta)
     \right) \, .
 \end{align}
Since $\delta \ll \delta_\text{max}$, inflation ends at 
\begin{align}
    \label{eq:psi_end}
    \psiend
    &\approx
    - \frac{2^{\frac{1}{4}} \sqrt{3}}{6} M \sqrt{\frac{M}{\mpl}}
    .
\end{align}
When expressed as  a function of $\psi$ the minimum of the potential is  at $\psi=-M$; the relative width of the inflationary saddle scales as $\sqrt{M/\mpl}$ and $\psi_{\text{begin}} \approx - \psiend$.

We can now evaluate the integral in \eqref{eq:N_integral}; if $\delta=0$ it is formally divergent for $\psin \ge 0$, and $N$ is apparently infinite.
For $\delta \ne 0$ \eqref{eq:N_integral} can be solved approximately by setting $V= V(0)$ and taking only the two lowest order terms in $V'$:
\begin{align}
    N
    &\approx
    \frac{1}{12} \frac{M^3}{\mpl^2}
    \int_{\psiend}^{\psin}
    \frac{%
        1
    }{%
        M^{2} \delta + \psi^{2}
    }
    d \psi
    \\
    &=
    \label{eq:N_integral_soln}
    \begin{cases}
        \frac{1}{12} \left(\frac{M}{\mpl}\right)^2
        \frac{1}{\sqrt{\delta}}
        \arctan\left( \frac{1}{\sqrt{\delta}} \frac{\psi}{M} \right)
        \bigg|_{\psiend}^{\psin}
        & \delta \ne 0
        \\
        - \frac{1}{12} \frac{M^{3}}{\mpl^2} \frac{1}{\psi} \bigg|_{\psiend}^{\psin}
        & \delta = 0
    \end{cases}
    \,.
\end{align}
These expressions apparently disagree in the limit $\delta\to0$ but the identity $\arctan(x) = -\arctan(1/x) - \pi/2$ makes their overlap manifest when $\psin$ has the same sign as $\psiend$:
\begin{align}
    N
    &\approx
    -
    \frac{1}{12} \left(\frac{M}{\mpl}\right)^2
    \frac{1}{\sqrt{\delta}}
    \arctan\left( \sqrt{\delta} \frac{M}{\psi} \right)
    \bigg|_{\psiend}^{\psin}
    \\
    &\approx
    - \frac{M^{3}}{12 M_{pl}^{2} \psi}
    + \frac{\delta}{36} \frac{M^{5}}{\mpl^{2} \psi^{3}}
        \qquad \text{for} \qquad 0 \leq \frac{\sqrt{\delta} M}{\psin} \ll 1
    \,.
\end{align}
 Consequently, when $\delta \rightarrow 0$ and $M < \mpl$ the field value at the pivot scale is approximately
\begin{equation}
    \label{eq:psi_n_approx}
    \psin
    \approx
    - \frac{1}{12} \frac{M^3}{\mpl^2} \frac{1}{N}
    .
\end{equation}

We need $\delta \ll 1$ for successful inflation but equation~(\ref{eq:psi_n_approx}) is valid only when $\delta$ is smaller than the bound in  eq.~\eqref{eq:delta_max2}:
as this bound becomes saturated the inflationary phase is relatively short and the pivot scales leaves the horizon when $\psi_N>0$. In this case $\eta$  changes sign during the observationally relevant phase of inflation and when $\eta$ is positive  $n_s$ can exceed unity. However, for fixed  $N$  there is a lower limit on $n_s$. This behaviour can be deduced from a slow roll analysis, and is apparent in the plots of $n_s$ and $r$ shown in Ref.~\cite{Bird:2008cp} for the spectral parameters 60 $e$-folds before the end of inflation. Similar behaviour is also seen in the related model studied in Ref.~\cite{BuenoSanchez:2006rze}; for a given $r$ there is sharp lower bound on $n_s$, and no clear upper bound.

The matching equation connects present-day scales to the  inflationary era \cite{Liddle:2003as,Adshead:2008vn,Adshead:2010mc}; assuming instant thermalization after inflation it is
\begin{align}
    N
    &=
    56.12 + \frac{1}{4} \log \frac{2}{3}
    + \log\left(\frac{V_N^{1/4}}{V_\text{end}^{1/4}}\right)
    + \log\left(\frac{V_N^{1/4}}{10^{16} \text{~GeV}}\right)
    \\
    \label{eq:matchin_equation_with_subs}
    &\approx
    61 + \frac{1}{4} \log \lambda
    + \frac{1}{4}
    \log\left(
        \frac{%
            \bigl[ M^4 + 12 M^3 \delta \psin]^{1/2}
        }{%
            \mpl
            \bigl[ M^4 + 12 M^3 \delta \psiend]^{1/4}
        }
    \right)
    .
\end{align}

\begin{figure}[tbp]
    \centering
    \includegraphics[width=0.95\linewidth, draft=false]{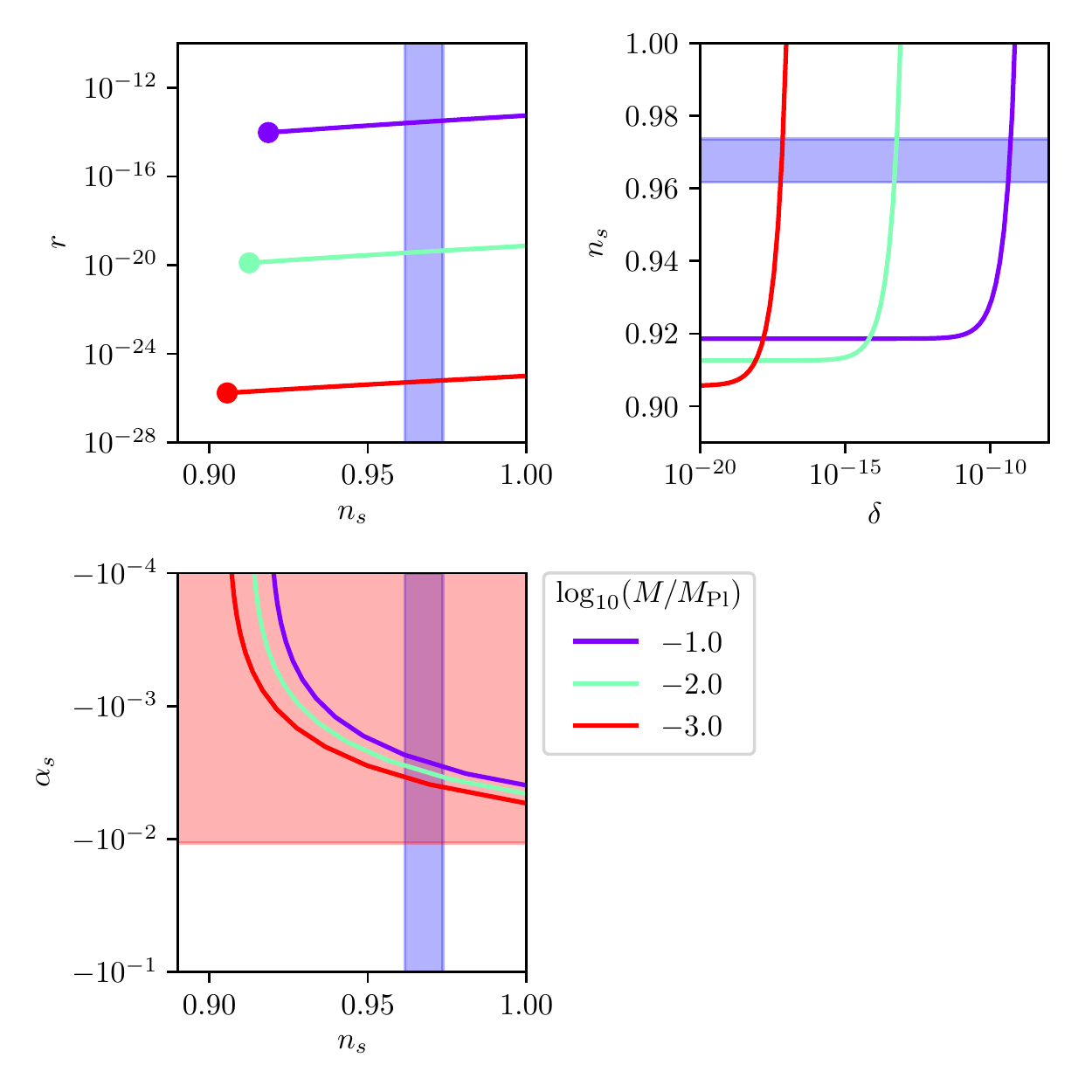}
    \caption{%
        Spectral parameters for three  values of $M$,  varying $\delta$ across its permissible range. The values of $\lambda$ and $N$ are fixed by simultaneously matching the amplitude to the observed value of $A_s$ and solving the matching equation assuming instant reheating. 
        We include indicative bounds on $n_s$ (blue) and $\alpha_s$ (red); 68\% confidence level results from Planck  \cite{Ade:2015lrj}.
        The points at the ends of the lines in the top left figure denote $\delta=0$.
    }
    \label{fig:spectrum_N_matching}
\end{figure}

Finally, the parameter $\lambda$ is fixed by the amplitude of the scalar perturbations:
\begin{align}
    \label{eq:scalar_amplitude}
    A_s
    &=
    \frac{1}{12 \pi^2 \mpl^6} \frac{V^3}{V'^2}
    =
    \frac{\lambda}{20736 \pi^{2}}
    \frac{
        M^7 \left(M + 12 \psin \delta \right)^{3}
    }{
        \mpl^{6} \left(M^{2} \delta + \psin^{2} \right)^{2}
    } \, .
\end{align}
The value of $A_s$ is  well-measured  \cite{Ade:2015xua}; in our numerical examples we take  $A_s = 2.2 \times 10^{-9}$. In the limit  $\delta \rightarrow 0$ equations~\eqref{eq:psi_n_approx} and \eqref{eq:scalar_amplitude}  give an expectation for the height of the potential:
\begin{equation}
    \lambda
    =
    \pi^2 A_s \left(\frac{M}{\mpl}\right)^2 \frac{1}{N^4} \, .
    \label{eq:lambda_small_field}
\end{equation}

In Figure~\ref{fig:spectrum_N_matching} we show  spectral parameters as a  function of $\delta$ computed self-consistently from the matching equation along with current bounds on these parameters.\footnote{We do not plot joint probability distributions on $r$, $n_s$ and $\alpha_s$; as we will see later, $\alpha_s$ may be strongly scale dependent for some parameter choices and direct observational constraints on this potential will be obtained in a forthcoming analysis.}
For any sub-Planckian $M$ the scalar-tensor ratio $r$ is very small, with $r\sim 10^{-14}$ for $M=0.1 \mpl$.
Moreover, when $\delta  \rightarrow 0$,  $n_s$ is below the  observationally allowed range, $n_s \lesssim 0.93$ vs $n_s = 0.9677\pm0.0060$.
However, $n_s$ increases as  $\delta$ approaches the bound in equation~\eqref{eq:delta_max2} and so this model is consistent with current data.

\begin{figure}[]
    \centering
    \includegraphics[width=0.95\linewidth, draft=false]{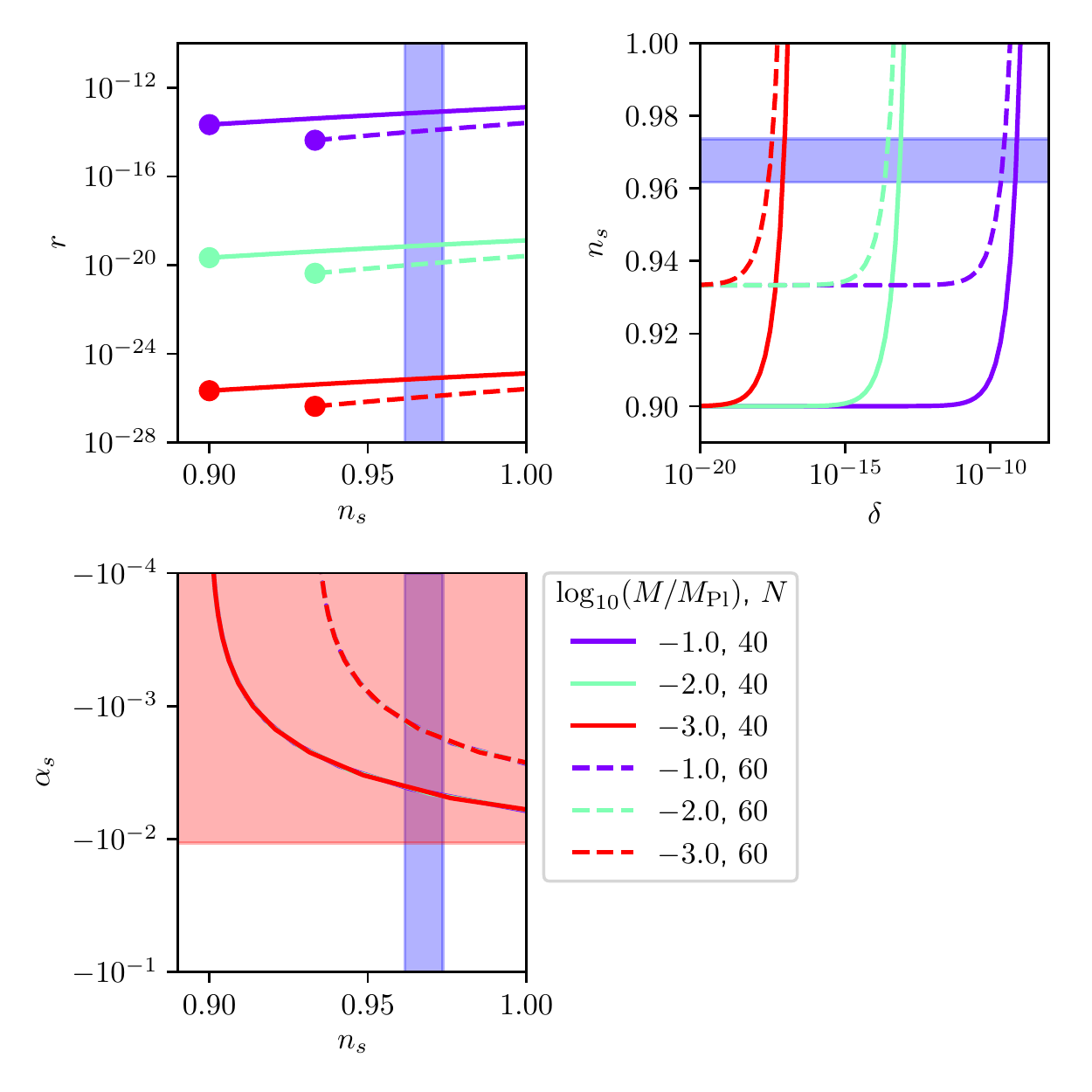}
    \caption{%
        We plot spectral parameters for three  values of $M$ and two values for $N$, while varying $\delta$ across its permissible range.
        As in Figure~\ref{fig:spectrum_N_matching} we have plotted current constraints on $n_s$ (blue) and $\alpha_s$ (red), and $\delta$ varies along each of the curves.
       Note that the curves  for $\alpha_s$ vs $n_s$ coincide.
    }
    \label{fig:spectrum_N_fixed}
\end{figure}

Interestingly, $|\alpha_s|$ also increases with $\delta$, providing leverage to test the model as constraints on the spectrum improve.
CMB-S4 in the microwave background ``roadmap'' aims to measure the running with an uncertainty of $0.002$ to $0.003$ \cite{Abazajian:2016yjj,Munoz:2016owz}, while the Square Kilometer Array (SKA) may reduce the 1-$\sigma$ bounds on $\alpha_s$ to $0.0018$ (SKA1) or $0.00092$ (SKA2) \cite{Pritchard:2015fia}; combining data from CMB-S4 and SKA2 would put even tighter constraints on $\alpha_s$ \cite{Munoz:2016owz}.
Consequently, planned future observations could conclusively falsify this model.

These conclusions would not change significantly if we drop the assumption of instant thermalization; Figure~\ref{fig:spectrum_N_fixed} shows the spectral parameters as a function of $N$.
In the $\delta =0$  limit, increasing $N$ pushes  $n_s$ closer to the observationally allowed range.
Physically, this would require a post-inflationary phase where the equation of state $w > 1/3$, which is technically feasible but has little  motivation.
Even then the $\delta =0$ limit cannot easily be made consistent with the data.
Conversely, if thermalization is preceded by a substantial matter dominated phase, $n_s$ is moved further away from unity.
Likewise, the running  $\alpha_s$  depends on $N$,  but $\left| \alpha_s \right|$ remains above $10^{-3}$ for any reasonable configuration.

This analysis   takes place in the slow roll limit.
This  amounts to a constraint on the initial conditions since, if the inflaton arrives at the plateau with a significant kinetic energy, it will experience ``ultra slow-roll'' and ``overshoot'' before inflation can commence \cite{Itzhaki:2007nk,Dimopoulos:2017ged}.
To show this we make the simplifying assumptions that $\ddot{\phi} + 3H \dot{\phi} \approx 0$ and $\dot{\phi}^2 \approx V$ as the field-point approaches the inflationary plateau; the latter condition is equivalent to requiring that $\rho + 3p \approx 0$, which is the condition for the onset of inflation \cite{Lidsey:1995np}.
In this limit, the field velocity is approximately $\dot{\phi}_0 \exp{(-3 H t)}$ where $\dot{\phi}_0$ is the velocity at a given initial time.
Via equation~(\ref{eq:psi_end_sq}) the total width of the inflationary plateau is roughly $M \sqrt{M/\mpl}$.
Putting all this together (and recalling that $H \sim \sqrt{V}/\mpl$) we see that $3H\Delta t \sim (M/\mpl)^{3/2}$ so for $M<\mpl$ the field velocity will not change dramatically as it evolves through the region containing the inflection point.
Consequently, inflation cannot commence unless $\dot{\phi}^2 \ll V(\phi)$ initially.
Imposing this condition is strong  tuning unless the model is embedded inside a larger dynamical system in which this situation occurs naturally -- a stipulation that undercuts any claim to simplicity.

\section{Large Field Inflation}

We now turn our attention to the large field limit.  As previously, there is no minimum other than the origin if $g \le 2 \sqrt{\lambda} m$. 
When $M\lesssim \mpl$ and $\phi \sim M$, inflation requires $\delta \ll 1$; when $M \gtrsim \mpl$ and $\phi \gtrsim \mpl$ inflation will occur whether or not the potential possesses an inflection point.  Moreover, negative values of $g$ are consistent with inflation in this regime; in that case all nontrivial derivatives of the potential are positive.
We will find it convenient to define $g= 2 \sqrt{\lambda} m \Delta$ and again use $M = g/2\lambda$ to write
\begin{align}
    \label{eq:}
    m &= \sqrt{\lambda} M \, ,
    \\
    g &= 2 \lambda M \Delta \, .
\end{align}
This allows us to express $V(\phi)$ as
\begin{equation}
    \label{eq:potential_large_field}
    V(\phi)
    =
    \lambda \left( \frac{M^2}{2} \phi^2 - \frac{2}{3}\Delta M \phi^3 + \frac{1}{4}\phi^4 \right)
    \, .
\end{equation}
and we have not dropped any terms from the potential.
The  contribution of the cubic term,  $\frac{2}{3}\Delta M \phi^3  / V(\phi)$, is maximized at $\phi=\sqrt{2} M$ so the transition from  a quadratic to quartic potential occurs when $\phi \sim M $ even if $V'$ increases monotonically.
In the  near-saddle point limit $\Delta \sim 1/\sqrt{1+\delta}$, where $\delta$ is the expansion parameter from the small field case.

Large field inflation occurs for any value of $\Delta$, although $\Delta \le 1$ is needed to avoid a trapping potential.
As in the small field case, $\lambda$ sets the amplitude of the perturbations but does not influence the inflationary dynamics.
We restrict our attention to scenarios where the velocity is consistent with slow roll; additional possibilities arise if we allow transient velocities but have little physical motivation. 
The $\{M,\Delta\}$ parameter space contains a number of distinguishable large field scenarios, which we now enumerate:

\begin{itemize}
\item {\em \noindent Effective Quadratic Potential} If $M\gg 10 \mpl$, the potential is effectively quadratic throughout the cosmologically relevant phase of inflation.

\item {\em \noindent Effective Cubic Potential} If $\Delta \gg 1$ and $M \gtrsim {\cal{O}}(\text{few})\mpl$, the potential is effectively cubic during the cosmologically relevant phase of inflation.

\item {\em Effective Quartic Inflation} If $M \lesssim \mpl$ the  potential is effectively quartic for $\phi \gtrsim \mpl $ and the resulting spectrum will be incompatible with the data \cite{Peiris:2003ff,Adam:2015rua,Ade:2015xua,Ade:2015lrj}.

\item {\em Exact Saddle Point: Eternal Inflation} Here $\Delta = 1$ and $M \gtrsim 2\mpl$. If $\phi$ is initially larger than $M$, the field dynamics have an attractor solution with $\phi\rightarrow M$ as $t \rightarrow \infty$ and inflation never ends. In the slow roll limit,
\begin{eqnarray}
&3 H \dot{\psi} + \lambda M \psi^2 \approx 0 & \label{hpsi}\\
\Rightarrow &  \psi \approx \frac{M}{2 \mpl \sqrt{\lambda}} \frac{1}{t}
\end{eqnarray}
 where again $\phi = \psi+M$ and $H$ is effectively constant. The neglected terms in equation~(\ref{hpsi}) all scale as $t^{-2}$ or beyond as $\psi \rightarrow 0$ so this solution has a non-trivial basin of attraction establishing that slow roll remains valid even as $V'$ vanishes identically. Consequently this configuration  supports eternal inflation and  semiclassical evolution must begin with $\phi < M$, both from the perspective of the field configuration in the primordial universe and in  numerical treatments of the inflationary dynamics. 

\item {\em Large Field Saddle Point Inflation} The most complex possibility is a saddle point lying in the cosmologically relevant segment of the potential, e.g. with $M \sim {\cal{O}}(\text{few})\mpl$ and $0 \lesssim \Delta < 1$.   In the large-field regime, $V'/V$ is necessarily small but  is further suppressed in the vicinity of a (near)-saddle point. The perturbation amplitude scales as $V^3/V'$ so when $V'$ passes through a local minimum there is a corresponding feature in the spectrum. The height of the feature is correlated with its width (as a function of comoving wavenumber, $k$) since $N \sim V/V'$,  ``stretching'' the feature over a larger range of $k$-values if $V'$ approaches zero.   As explored below, this scenario  produces a range of outcomes including the counterintuitive possibility that the scalar amplitude is reduced relative to a comparable monomial case at the same energy density.

\item {\em Punctuated Inflation}
    If  $\mpl \lesssim M \lesssim 2\mpl$ and $\Delta \approx 1$ a punctuated inflationary scenario is possible \cite{Jain:2008dw,Jain:2009pm}, as illustrated in Figure~\ref{fig:traj}.
    These solutions   arise  when $\Delta$ is very close to unity and in the exact $\Delta=1$ limit.
    When $\mpl \lesssim M \lesssim 2\mpl$ the $\Delta = 1$ limit is far enough from perfect slow roll to evade the eternal inflation solution above.

\begin{figure}[p]
    \centering
    \includegraphics[width=0.9\linewidth, draft=false]{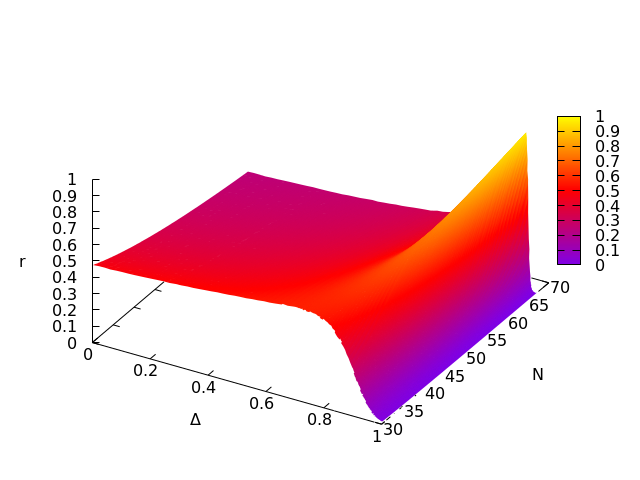}
    \includegraphics[width=0.9\linewidth, draft=false]{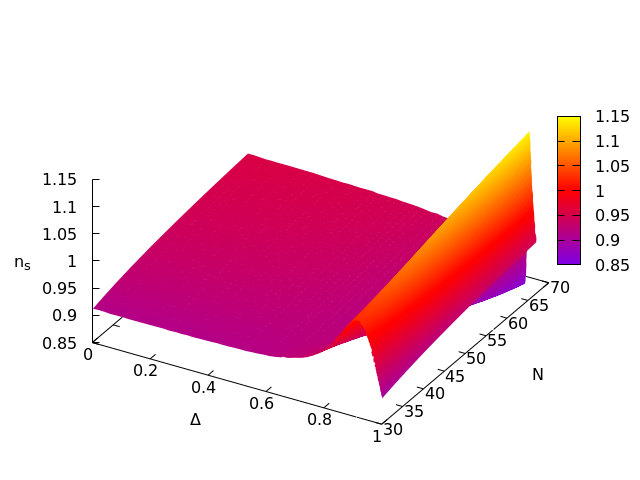}
    \caption{%
        Spectral parameters $r$ and $n_s$ as a function of $\Delta$ and $N$, the number of $e$-folds before the end of inflation. In each case $M=8\mpl$.
            }
    \label{fig:large-M-N_Delta-M010}
\end{figure}

\afterpage{\clearpage}

\item {\em Hilltop Inflation}  If $\Delta>1$ hilltop inflation results; hilltop inflation yields a relatively large value of $r$ unless the quadratic term in the Taylor expansion vanishes almost exactly \cite{Kinney:1995ki,Easther:2006qu}, a situation that cannot occur for this potential.   
 
 \end{itemize}

\begin{figure}[tbp]
    \centering
    \includegraphics[width=0.7\linewidth, draft=false]{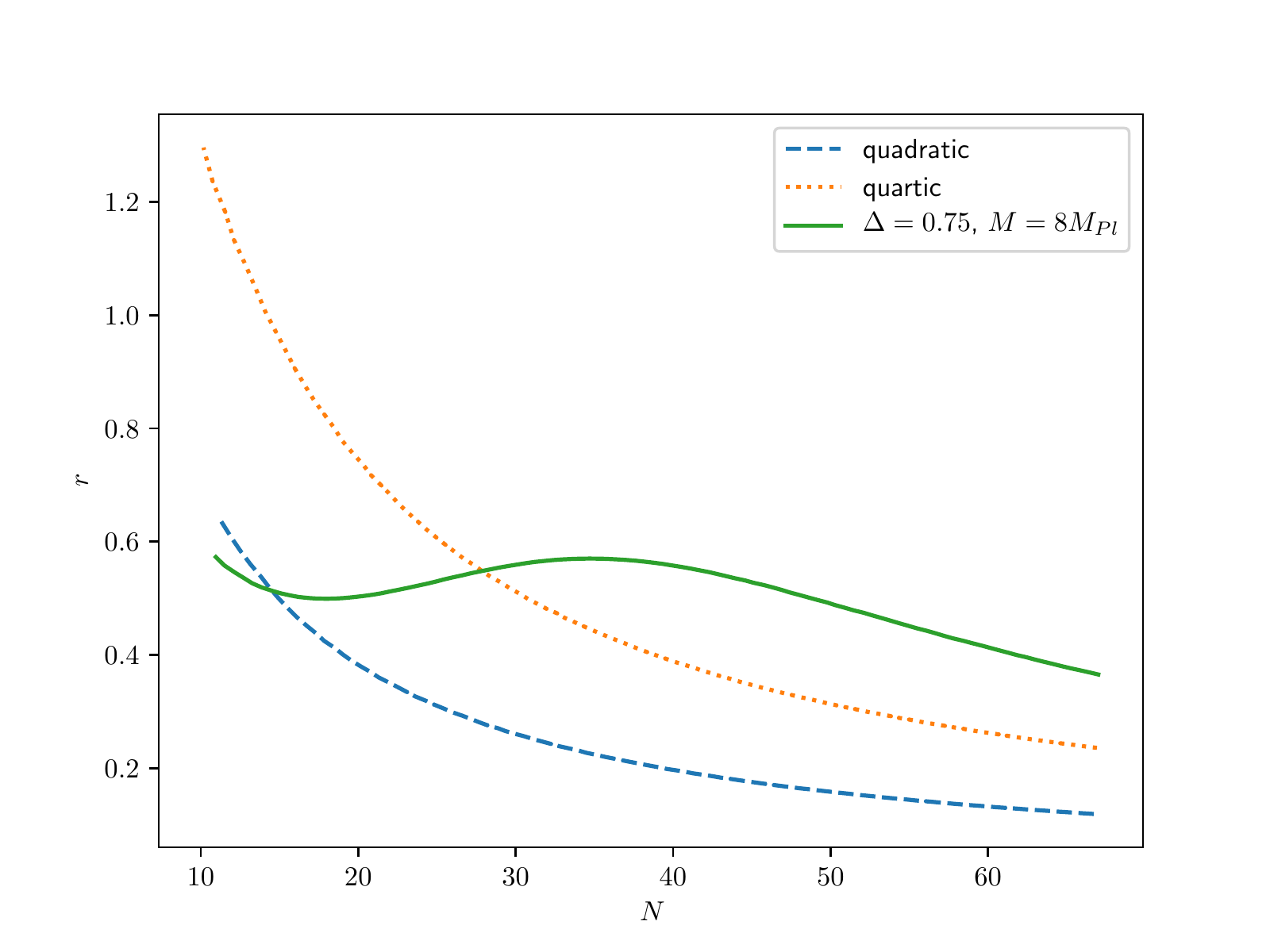}
    \caption{%
        The tensor to scalar ratio $r$ is plotted for three models, quadratic (blue), quartic (orange) and $M=8\mpl$ and $\Delta = 0.75$ (green); data computed with ModeCode assuming instant preheating with $\lambda$ chosen to reproduce the observed  density perturbation amplitude.       The tensor amplitude of the inflection point scenario can exceed that of either monomial scenario.
    }
    \label{fig:r_values}
\end{figure}

\begin{figure}[tbp]
    \centering
    \includegraphics[width=0.7\linewidth, draft=false]{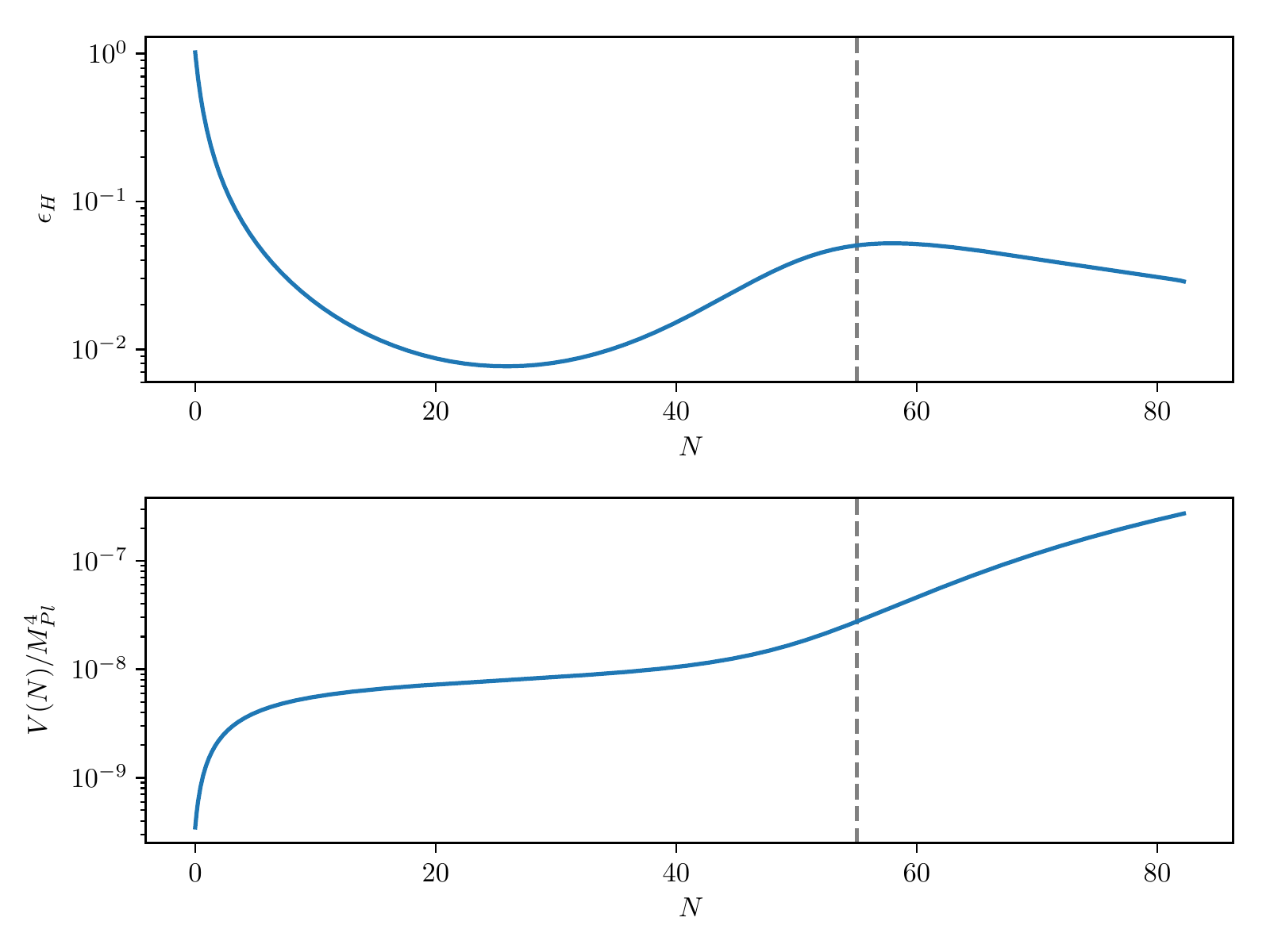}
    \caption{
        Trajectory for a set of parameters ($M=8 \mpl$, $\Delta=0.935$, $\lambda = 1.6 \times 10^{-11}$) yielding a large tensor-scalar ratio $r = 0.8$ when $N$ is fixed at $55$.
    }
    \label{fig:large-r-traj}
\end{figure}

\afterpage{\clearpage}

Some of these scenarios are limited to a narrow region of the $\{M,\Delta\}$ parameter space but unlike the small-field regime there is no need for  tuning to ensure that inflation takes place.   In most cases the power spectrum can be well-understood in the slow-roll limit, but there are regions of parameter space for which this approach would be inadequate and we have implemented both the small field and large field scenarios in ModeCode \cite{Price:2014xpa}, which we use to generate the results shown here.\footnote{ModeCode was modified to add the potential~\eqref{eq:potential_large_field}; the initial conditions must be chosen with some finesse and the stopping condition must account for the possibility of punctuated inflation.}

\begin{figure}[tpb]
    \centering
    \includegraphics[width=0.95\linewidth, draft=false]{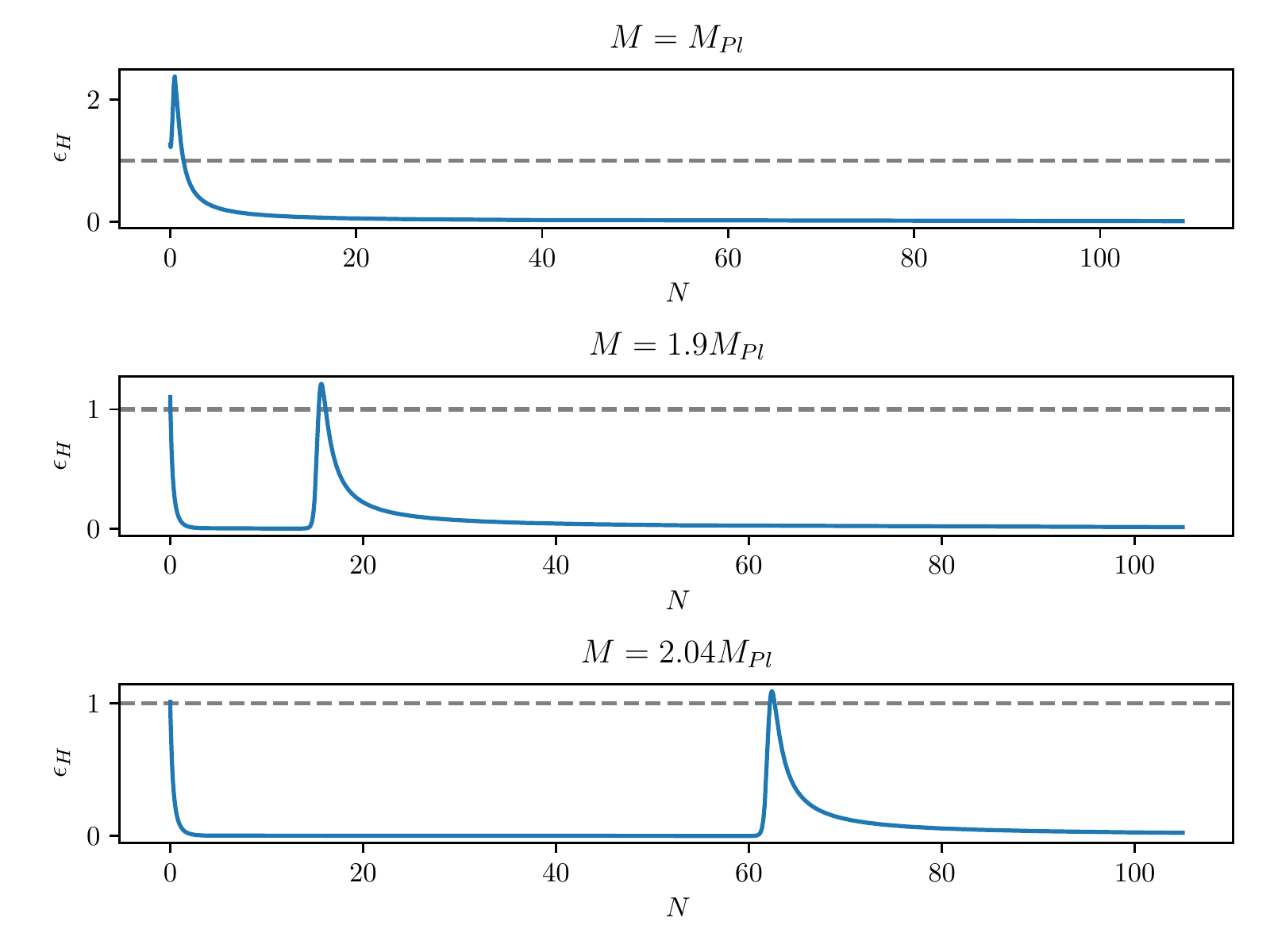}
    \caption{
    The parameter $\epsilon_{H}$, $N$ $e$-folds before the end of inflation; $\epsilon_{H} < 1$ if the expansion of the universe is accelerating.
    Each plot is with $\Delta = 0.9999$.
    For small $M$, the inflaton rolls through the inflection point without inflation  resuming; inflation is effectively quartic.
    For $1.06 \mpl \lesssim M \lesssim 2.04\mpl$ inflation pauses briefly. For larger $M$ all of the observationally relevant portion of inflation occurs in the vicinity of the inflection point.
    }
    \label{fig:traj}
\end{figure}

\begin{figure}[tpb]
    \centering
    \includegraphics[width=0.95\linewidth, draft=false]{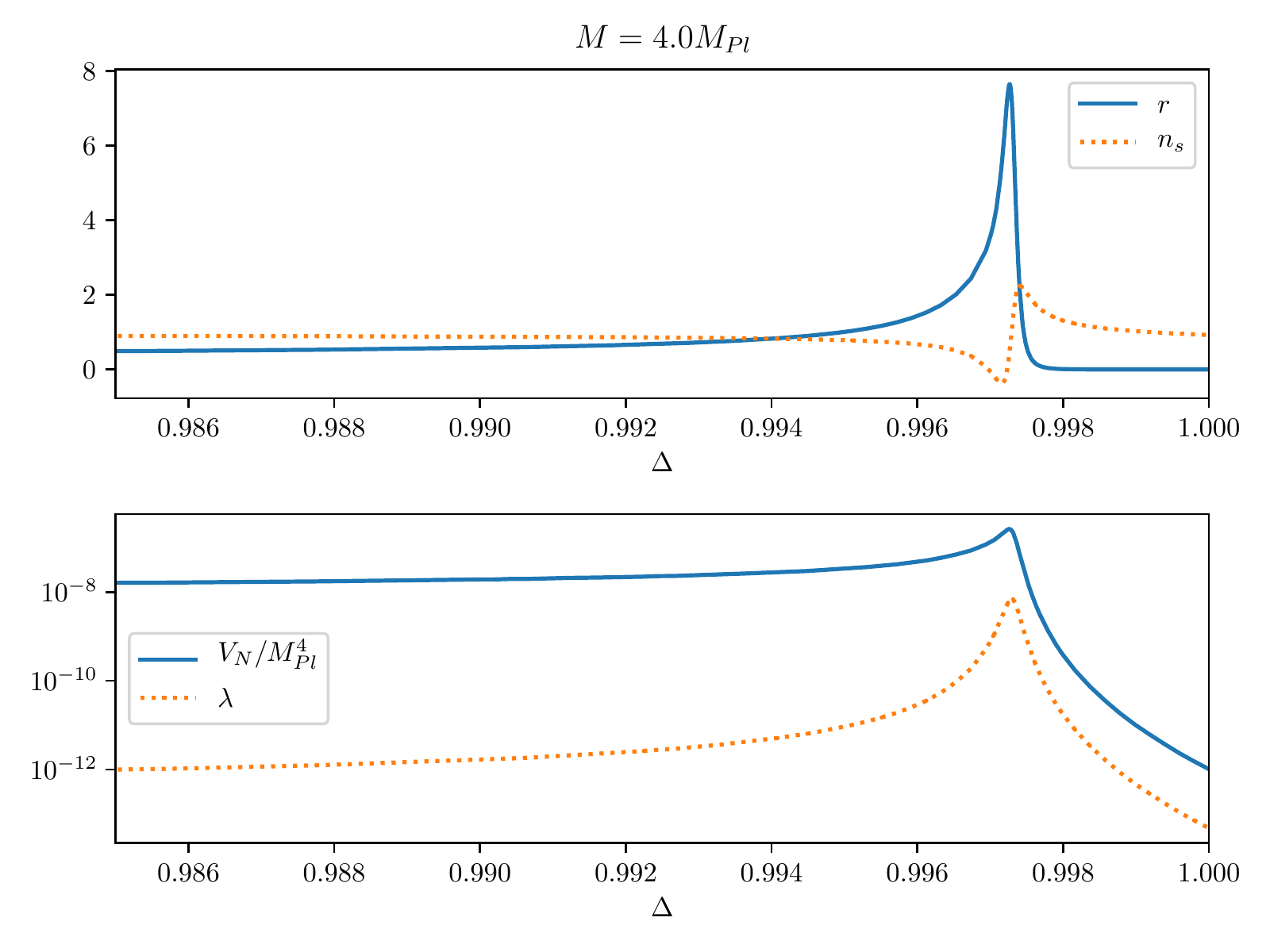}
    \caption{
        The top plot is of $n_s$ and $r$ as functions of $\Delta$ when $M = 4 \mpl$.
        The lower plot shows that large values of $r$ are accompanied by a higher energy scale at horizon crossing as $\lambda$ must increase to match the observed perturbation amplitude when $V/V'$ decreases.
    }
    \label{fig:large-M-Delta-cut-M=004}
\end{figure}

The most interesting scenarios from our perspective are those with  $0\le\Delta\le1$ and $\mpl<M \lesssim 20 \mpl$.
Figure~\ref{fig:large-M-N_Delta-M010} shows the astrophysical observables $n_s$ and $r$ as a function of $\Delta$ and $N$, the number of remaining e-folds.
These plots show that $n_s$ varies significantly thanks to the ``feature'' in the potential. However, for these scenarios $r$ can be  counter-intuitively large:  the cubic term reduces both $V$ and $V'$ but the impact on $V$ is larger than that on $V'$, increasing $\epsilon$ and thus $r$ relative to values seen with monomial potentials. A representative scenario is shown in Figure~\ref{fig:r_values}.
Given the observational constraints on $r$ this significantly reduces the area of the $\{M,\Delta\}$ plane that yields observables  compatible with current data, and runs contrary to the naive expectation that a flatter potential necessarily leads to a lower value of $r$.

As illustrated by Figure~\ref{fig:large-r-traj},   models with a large $r$ are those where the inflaton is approaching the plateau in the potential as astrophysical modes leave the horizon. In these cases $\epsilon$ and thus $r$ is significantly scale dependent.  Models with large $r$  are now primarily of academic interest, but note that the tensor amplitude is proportional to $V$ and thus slowly decreasing; the rapidly increasing scalar amplitude  drives the scale-dependence  of $r$. The punctuated inflation scenarios can be understood as extreme versions of this situation. Recalling that
\begin{equation} 
    \epsilon_H = \frac{1}{2} \frac{1}{\mpl^2} \left(\frac{d \phi}{d N}\right)^2
\end{equation}
is exactly unity when accelerated expansion ends, Figure~\ref{fig:traj} shows several punctuated scenarios in which inflation pauses briefly before resuming. Interestingly, these scenarios only exist when $M\gtrsim\mpl$. For smaller values of $M$ the field  rolls past the plateau before inflation can resume, further  demonstrating the ``ultra slow-roll'' and ``overshoot problem'' faced by the small field models.

For any model the specific value of $\lambda$ can be obtained by self-consistently solving the matching equation for $N$ and matching to the observed spectral amplitude. Given this constraint, a large value of $r$ implies that the energy density during inflation  is  higher than that in the low $r$ configurations.  The  height of the potential at the pivot scale is shown as a function of $\Delta$ in Figure~\ref{fig:large-M-Delta-cut-M=004}, assuming instant thermalisation.

\begin{figure}[p]
    \centering
    \includegraphics[width=0.8\linewidth, draft=false]{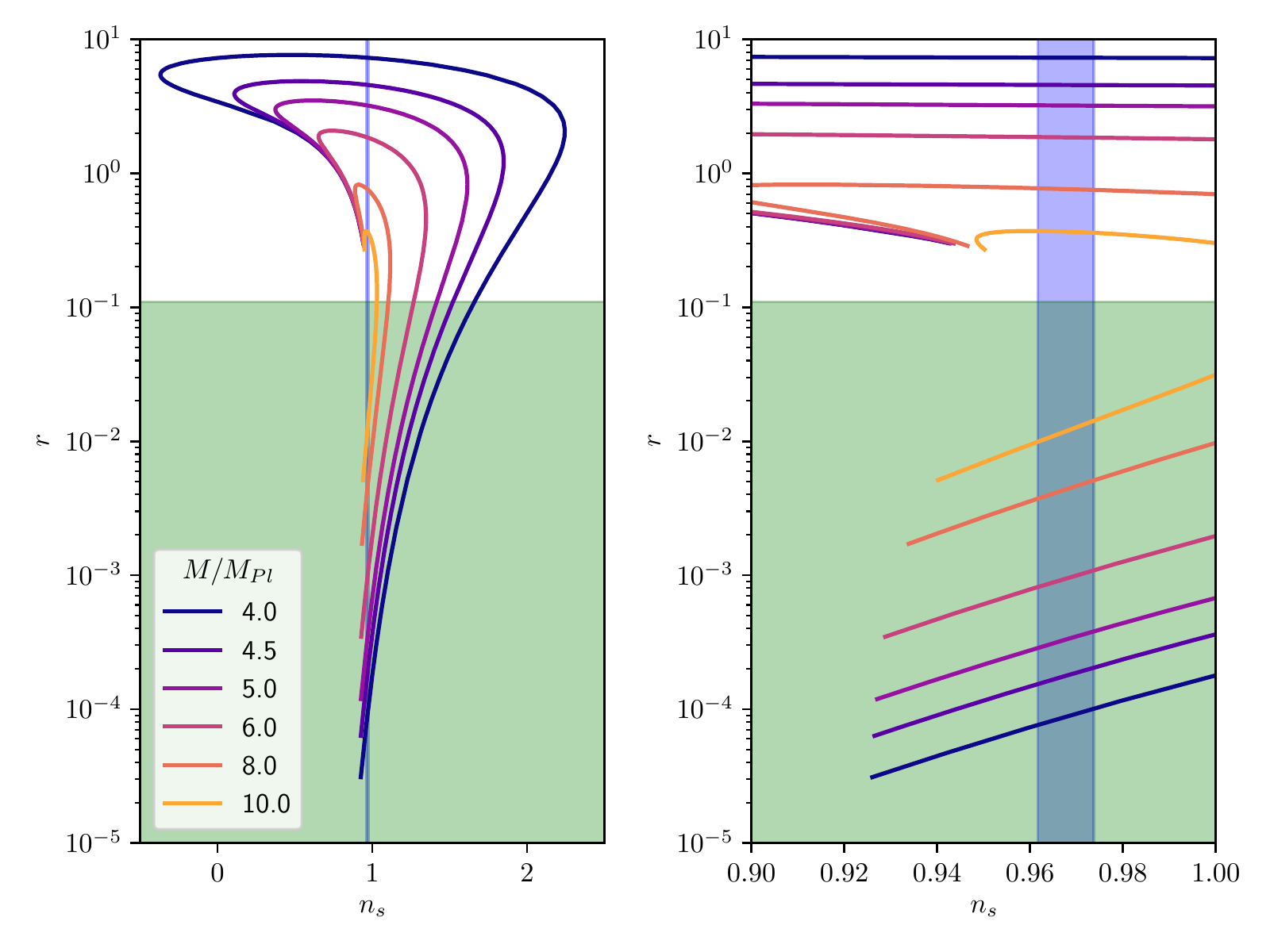}
    \includegraphics[width=0.99\linewidth, draft=false]{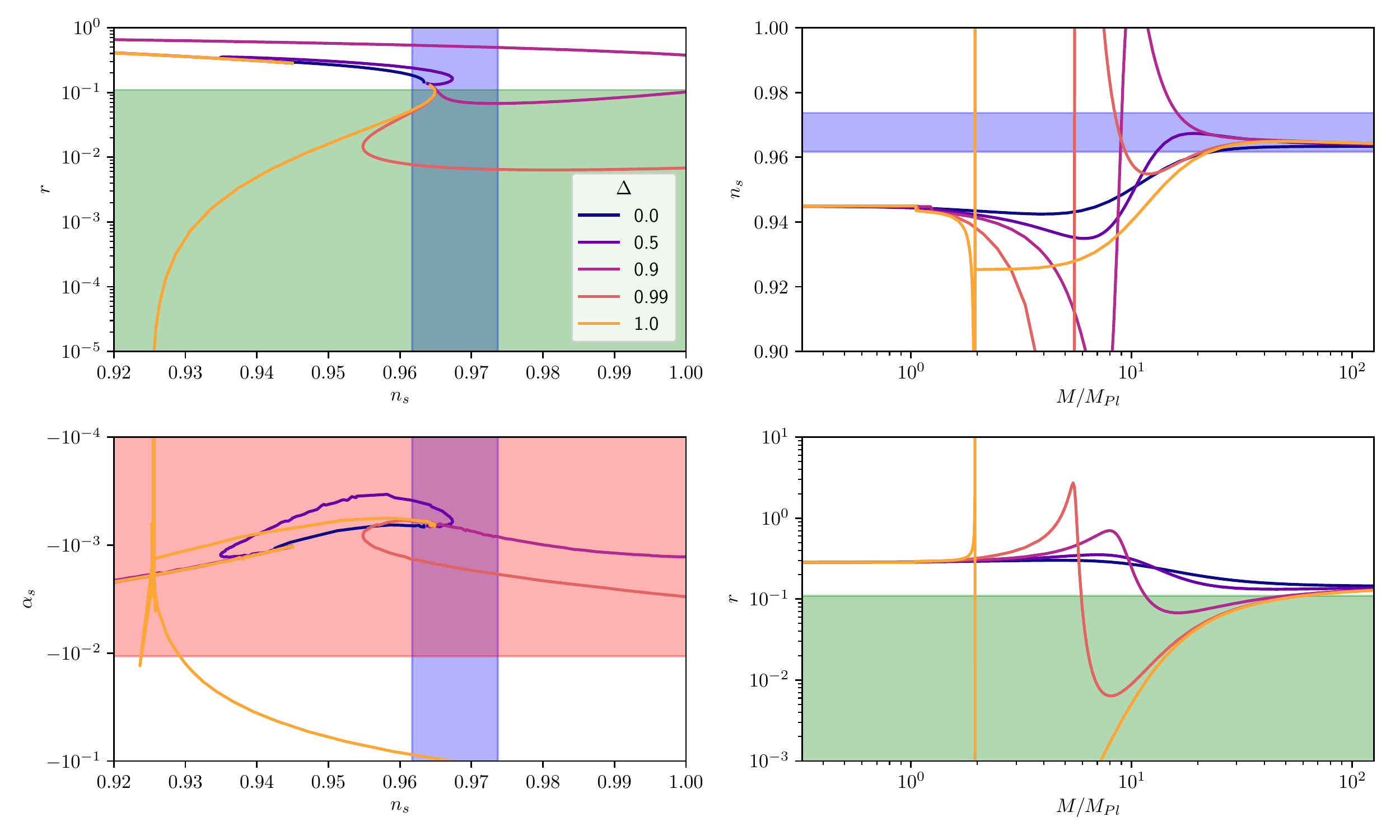}
    \caption{%
        In the top two panels $\Delta$ runs from $[0, 1]$ values of $n_s$ and $r$ consistent with current data are shown in blue and green.   The lower two rows show trajectories for fixed $\Delta$ and  $0.1 \mpl < M < 100 \mpl$.
        The inflationary phase is at large $\phi$ and we assume instant preheating with $\lambda$ chosen to reproduce the observed  amplitude at the pivot.
        The small discontinuity at $M \sim ~1$ corresponds to the transition between the first and second cases of Figure~\ref{fig:traj}.
    }
    \label{fig:constraints}
            \label{fig:large-M-vary-Delta-spectrum-MultiModeCode}
\end{figure}

In Figure~\ref{fig:large-M-vary-Delta-spectrum-MultiModeCode} we show $n_s$ and $r$ measured at the pivot scale for a range of $M$ and $\Delta$.
The limit $\Delta\rightarrow1$ is  a saddle point model with the field  rolling away from a plateau toward the global minimum; in these cases $r$ can be very small and $n_s$ lies below the observationally permitted range.
A preference for relatively large values of the running (as compared to typical single-term potentials \cite{Adshead:2010mc}) is clear, and we we also see that models in which the running is small typically have large values of $r$, an observation we explore in  detail in the following Section.

\afterpage{\clearpage}

\section{Fine-tuning, Priors and Testability}

The preceding Sections catalog the wide range of inflationary dynamics associated with the quartic polynomial potential.  We now use these analyses to specify priors for these scenarios  that  will facilitate  parameter estimation and model selection calculations.  In principle, one can work with a generic quartic potential with priors specified in terms of the slow roll parameters\footnote{As done in \cite{Aslanyan:2015hmi} for potential slow roll parameters, or \cite{Norena:2012rs} for a three parameter Hubble Slow Roll analysis; in the latter case the potential is not strictly a quartic polynomial.} but here we build on the analysis of the previous sections to parametrize these inflationary scenarios.\footnote{In the context of  Bayesian model selection, two ``models'' with the same algebraic  specification and which differ only in the joint distribution from which their parameters are drawn and are viewed as distinct scenarios, since the evidence integral will be weighted differently for each case.}  In particular, these priors will facilitate  parameter estimation and model selection calculations, see e.g. \cite{Peiris:2006ug,Mortonson:2010er,Easther:2011yq,Planck:2013jfk,Martin:2013nzq,Ade:2015lrj}, for these scenarios. 

Specifying the   distributions from which the parameters are drawn is a necessarily qualitative process.
However, the choices made when constructing these distributions weight the evidence  integrals (via equation~\eqref{eq:evidence}) and  this issue can be particularly pressing for multiparameter models. Approaches to specifying ``maximal entropy'' priors for inflationary models were examined carefully in Ref.~\cite{Easther:2011yq}.
Here the overall amplitude of the spectrum is set by a multiplicative parameter in front of the potential in both the large and small field regimes. In the small field case we can estimate  the likely value of  $\lambda$ (via equation~(\ref{eq:lambda_small_field})) but in the large field case we have to allow a large enough range of $\lambda$ to account for scenarios like those in Figure~\ref{fig:large-M-Delta-cut-M=004}.
We also apply a further cut --  any parameter choice for which the spectral amplitude does not satisfy $10^{-11} \le A_s \le 10^{-7}$ is excluded from the prior volume.
This will have no impact on the posterior distributions but protects against a ``volume effect'' when calculating  evidence \cite{Easther:2011yq}, a task we will pursue in a followup publication.

\subsection{Small Field}

To handle the small field case we first make the following parameterization 
\begin{eqnarray}
\delta &=&  \frac{\tilde{\delta}}{288} \left( \frac{M}{\mpl} \right)^4\\
\lambda &=& 10^{-9} \pi^2 \tilde{\lambda}   \left(\frac{M}{\mpl}\right)^2 \frac{1}{50^4} \, .
\end{eqnarray}
Inflationary solutions in this regime need an almost-exact inflection point for which the potential is described by equation~(\ref{potential_psi}).
In this setting, $M$ is an unknown energy scale and by assumption $M\le\mpl$.
The lower bound $M_{\rm min}$ is less clear;  in principle it could be as low as the TeV scale,  the minimal energy at which we can reasonably expect to see new physics, but in this limit the required tuning would be extreme. 
\begin{itemize}
\item {\em Small field, log prior: }  If $M$ corresponds to an unknown scale it is appropriate to draw it from a logarithmic or Jeffries prior:
\begin{eqnarray}
&\log_{10}{\left(\frac{M_{\rm min}}{\mpl}\right)} \le \log_{10}{\left(\frac{M}{\mpl}\right)} \le 0  \,  ,& \nonumber \\
& 0 \le \tilde \delta \le  1  \,  , &\\
&-5 \le \log_{10}(\tilde \lambda) \le 5 \nonumber \,  .&
\end{eqnarray} 
\item {\em Small field, uniform prior: } If  $M$ is assumed to be associated with an intrinsically high scale it is self-consistently drawn from a uniform prior, or
\begin{eqnarray}
    &\frac{M_{\rm min}}{\mpl}  \le \frac{M}{\mpl} \le 1  \, ,&   \nonumber \\
& 0 \le \tilde \delta \le  1  \,  , &\\
&-5 \le \log_{10}(\tilde \lambda) \le 5  \nonumber \,  .&
\end{eqnarray}
\end{itemize}

Inflationary models typically require a ``small parameter'' to ensure that the perturbation amplitude matches observation.
However, this potential requires not one but two small parameters -- $\lambda$ to set the overall scale of the potential and  fix the perturbation amplitude, and $\delta$ to quantify the departure of the inflection point at $\psi=0$  from an exact saddle.
As we saw in Section~3, inflation will only occur when $\delta \lesssim (M/\mpl)^4$.
Given current constraints on $n_s$ the limit $\delta \equiv 0$ is excluded by the data, as illustrated by Figures~\ref{fig:spectrum_N_matching} and~\ref{fig:spectrum_N_fixed}.
Consequently, the posterior for $\tilde{\delta}$ will differ substantially from the prior. If a symmetry is responsible for generating the saddle point it must  be weakly broken  by higher order contributions,\footnote{We justified the choice of a 4-th order polynomial, in part, by appealing to renomalization requirements but the actual inflaton dynamics will be controlled by the semi-classical potential which includes all loop corrections to the tree-level action.} but these corrections must be exceptionally small to prevent the inflection point  phase from being completely destabilized. Moreover, beyond the  tuned parameters and the ``overshoot'' problem analysed in Section~3, models in which inflation is supported by a narrow range of field values typically need highly homogeneous initial field configurations \cite{Goldwirth:1989pr,East:2015ggf,Clough:2016ymm}, partially undermining the explanatory power of inflation.  These problems could be ameliorated if small field inflation is preceded by a tunnelling event   \cite{Allahverdi:2008bt} but extra structure would need to be added to the theory to permit this, undercutting any claim to simplicity.  We have not attempted to ``score'' the dynamical tunings when constructing the priors, but these are more pronounced at lower values of $M$ which are disfavoured in the uniform prior relative to the log prior. 
 
 \subsection{Large  Field}

We now consider the large field case for which relevant values of $M$ lie between $\mpl$ and (generously) $50 \mpl$ -- beyond this we are in the quadratic limit of the theory and increasing $M$ will have little impact on observables;  given this relatively limited range we draw $M$ from a uniform distribution.  Conversely, we can draw $|1-\Delta|$ from either a  logarithmic  or a uniform distribution, depending on whether we understand the inflection point as arising from a (near) symmetry or an accidental cancellation, respectively.  Further,  $\Delta-1$ can have either sign and these cases are physically distinct, since $\Delta>1$ yields a local maximum with a trapping potential for which the onset of inflation and  associated initial conditions problem differs significantly from the case where field values can be arbitrarily large.\footnote{In principle we could have also considered a small-field hilltop scenario but it cannot yield physically reasonable spectra.}
\begin{itemize}
\item {\em Large field, log prior: }
        This scenario is appropriate if it is assumed that a near-saddle point is required by the symmetries of some underlying theory.
\begin{eqnarray}
&\mpl \le M  \le 50 \mpl  \,  , &\nonumber \\
& -6 \le \log_{10}(|\Delta -1|)  \le  0  \,  , &\\
&-15 \le \log_{10}( \lambda) \le -5 \nonumber \,   .&
\end{eqnarray} 
\item {\em Large field, uniform prior: }  This  scenario is appropriate if the near-inflection point  arises from an ``accidental'' cancellation of terms.
\begin{eqnarray}
&\mpl \le M  \le 50 \mpl   \,  ,& \nonumber \\
& 0< |\Delta-1|  \le  2  \,  ,&\\
&-15 \le \log_{10}( \lambda) \le -5 \nonumber \,  .&
\end{eqnarray} 
\end{itemize}
The ranges of some free parameters (e.g. the upper bound on $M$) cannot be inferred from fundamental principles and so we set the endpoints such that further extending the range would not introduce new possible combinations of empirical observables.   

 \begin{figure}[tbp]
    \centering
 \includegraphics[width=0.95\linewidth, draft=false]{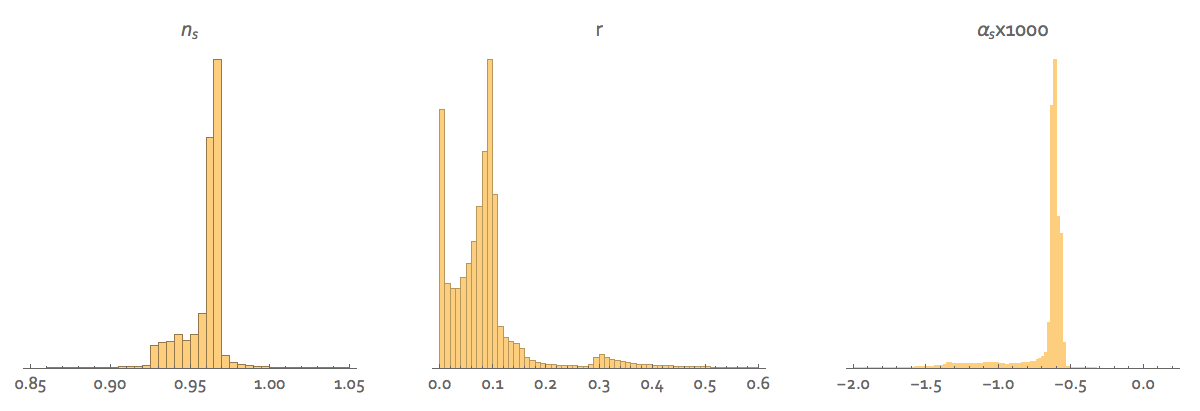}  
  \includegraphics[width=0.95\linewidth, draft=false]{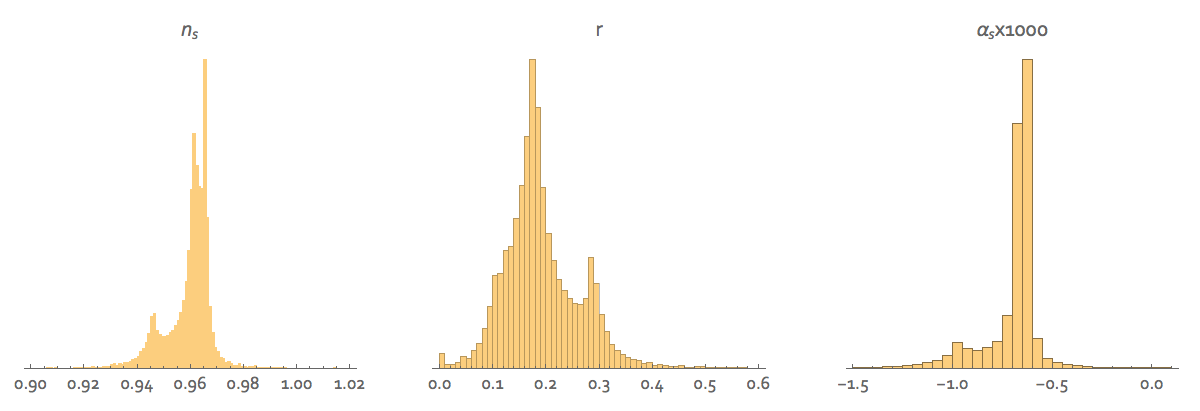}  
  \caption{%
The prior probabilities for the empirical spectral variables derived for the log  (top) and uniform hyperpriors (bottom) describing large field inflection point models.   }
\label{fig:derived_priors}
\end{figure}

 \begin{figure}[t]
  \centering
    \includegraphics[width=0.49 \linewidth, draft=false]{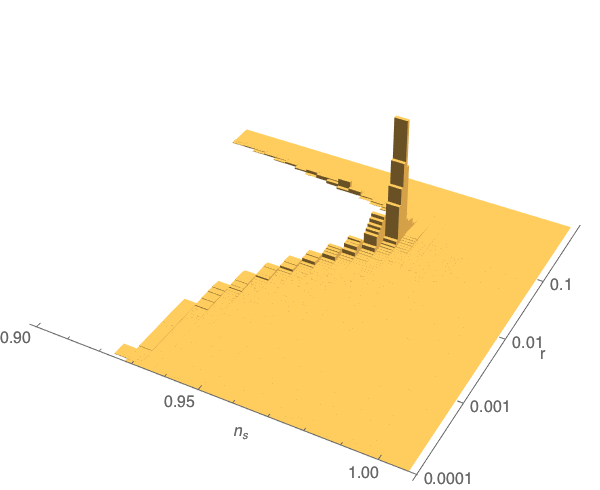}\\
   \includegraphics[width=0.49\linewidth, draft=false]{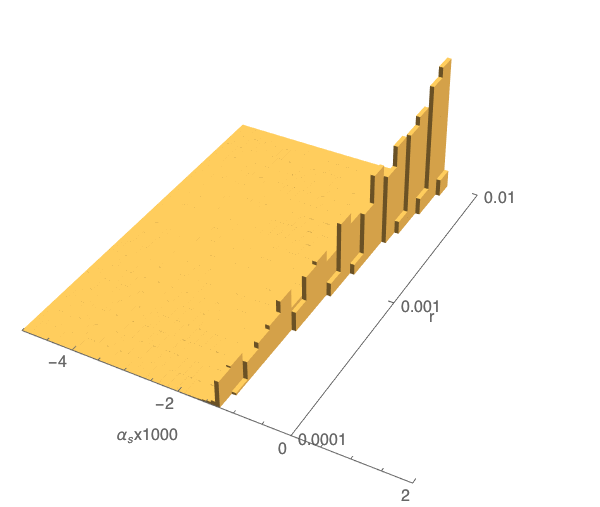}
      \includegraphics[width=0.49\linewidth, draft=false]{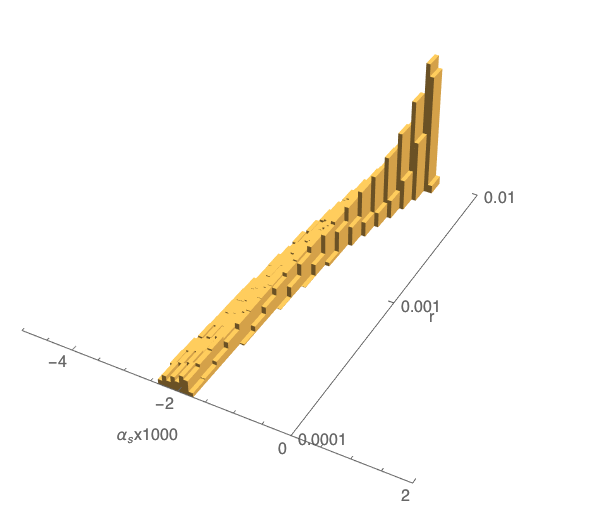}
  \caption{%
  Joint distributions for the large field, log prior inflection point scenario, measured  $N=55$ $e$-foldings before the end of inflation, with observables computed in the slow roll approximation. The top plot shows the joint  distribution of $r$ and $n_s$; the lower plots show $r$ and $\alpha_s$ -- the left hand plot  is restricted to parameter combinations for which $.9<n_s<1.1$ while the right hand plot shows only those with $0.95<n_s<0.97$.          }
                 \label{fig:joint_distro}
\end{figure}

These models can viewed as hyperpriors for hyperparameters $\Delta$ and $M$  in a Bayesian network that defines a generative model yielding $n_s$, $r$ and $\alpha_s$ \cite{Price:2015qqb}. If the running is large the scale dependence of $\alpha_s$ may also be nontrivial; however these scenarios are generically also those for which $r$ is much larger than observationally permitted. In what follows we will assume instant reheating and that the spectrum can be fully described by  $r$, $\alpha_s$ and $n_s$.  The resulting distributions of spectral variables for  large field inflection point scenarios are shown in Figure~\ref{fig:derived_priors} and they  are anything but uniform.  For both scenarios there is considerable support for models with $r<0.1$.

 Figure~\ref{fig:joint_distro}  shows joint distributions for the spectral parameters generated from the logarithmic large field prior. The degree of fine-tuning required to produce any given set of observables can be inferred from these plots, relative to the specified hyperprior.  This distribution is peaked at  $n\approx 0.963$, $r\approx0.14$, a pairing broadly consistent with quadratic inflation.   However,  the distribution of $r$  in Figure~\ref{fig:derived_priors}  reveals that there is an overall preference  for $r<0.1$ with the logarithmic prior. The distribution for $r$ derived from a uniform prior on $\Delta$ has a similar peak but favours larger values of $r$ relative to the logarithmic case. These results can be contrasted with the  tuning criterion advocated by Boyle, Steinhardt and Turok \cite{Boyle:2005ug} for which the ``least tuned'' region of the $\{n_s,r\}$ plane is larger than the peak found here, while ascribing a high level of tuning to regions of parameter space that are not strongly disfavoured by this Bayesian analysis.\footnote{This statement is, to some extent, dependent on the choice of upper bound on $M$ in the priors -- if the upper bound becomes arbitrarily large the joint distributions will become more peaked at the values associated with purely quadratic inflation. However, simply doubling this bound (to $100\mpl$) would not materially change our conclusions.}

A second noteworthy feature of the joint distributions  is that some combinations of parameters cannot be produced by any configuration of the 4-th order potential, so future observations could comprehensively falsify the overall scenario.  ``Excluded regions'' exist in both the $\{n_s,r\}$ and $\{\alpha_s,r\}$ planes.  For example, if $n_s$ and $r$ had been accurately measured to be $0.93$ and $0.1$ respectively all 4-th order potentials would be ruled out; however $n_s$ is now known to be larger than this limit.\footnote{The same structure in the $\{n_s,r\}$ plane is  visible in the plots of Refs~\cite{Boyle:2005ug,Bird:2008cp}.}  However, the permitted pairings of  $\{\alpha_s,r\}$ are restricted by limits on $n_s$ so we can update these priors in the light of observational evidence, as illustrated by Figure~\ref{fig:joint_distro}.

Looking again at Figure~\ref{fig:joint_distro} we see that small values of $r$ are correlated with relatively ``large'' values of $\alpha_s$, in contrast to most widely studied, simple models of single field inflation for which the running is typically $\alpha_s \sim -{\cal{O}}(\text{few}) 10^{-4}$ \cite{Adshead:2010mc}.
Likewise, the distributions for $\alpha$ shown in Figure~\ref{fig:derived_priors} peak at $|\alpha_s| < 10^{-3}$.
Projections suggest that SKA2 will be able to measure  $\alpha_s$ to a precision of $0.001$ \cite{Pritchard:2015fia} and Stage-IV CMB experiments likewise hope to measure $r$ to a precision of $0.001$ \cite[Table~6-2]{Abazajian:2016yjj}.
Consequently, even if future high-precision cosmological measurements only put tight upper bounds on  the running and the tensor amplitude this would suffice to rule out all possible inflationary scenarios built upon a single minimally coupled field with a 4-th order polynomial potential.\footnote{A similar correlation between $r$ and $\alpha_s$ is observed in Ref.~\cite{Boyle:2008ri} in the context of the Hubble Slow Roll approximation, in which the Hubble parameter $H$ is represented as a finite order polynomial in $\phi$.}

\begin{figure}[p]
    \centering
    \includegraphics[width=0.8\linewidth, draft=false]{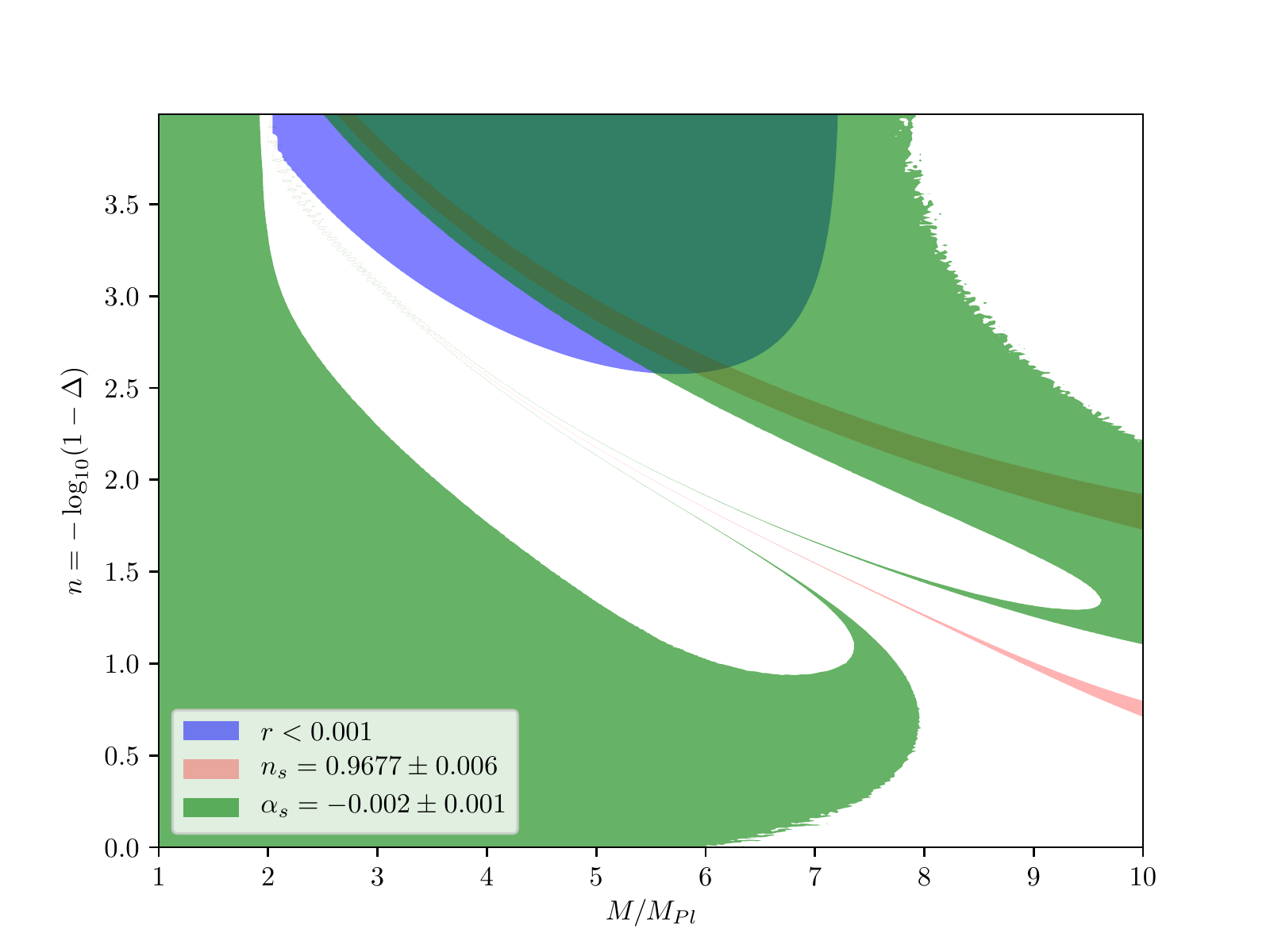}
   \includegraphics[width=0.8\linewidth, draft=false]{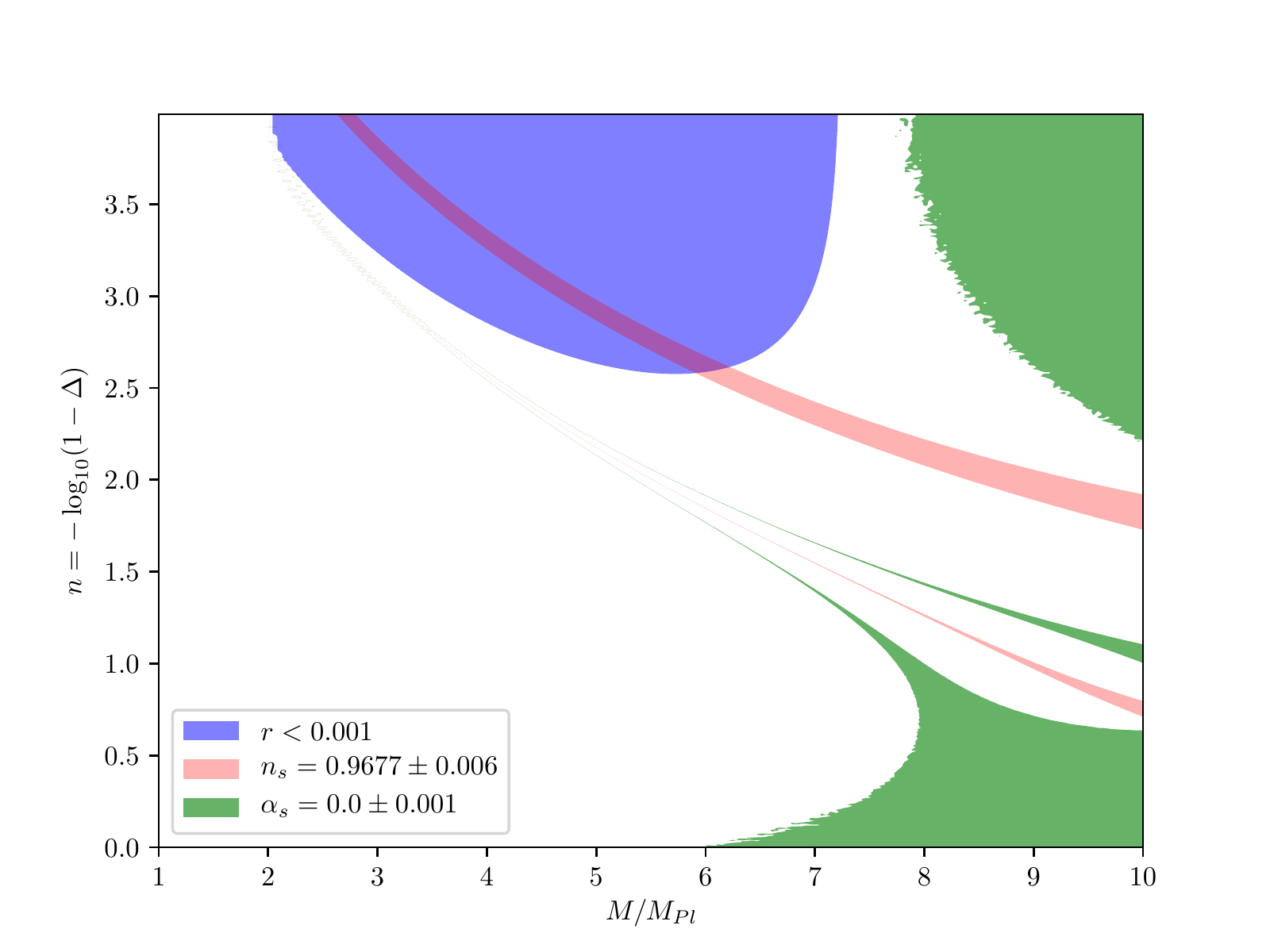}
   \caption{%
   Contours in the $M$-$\Delta$ plane corresponding to the assumption that $\delta_{\alpha_s} < 0.001$ and $r<0.001.$
        Measurements of this precision are consistent with  SKA2 \cite{Pritchard:2015fia} and Stage-IV CMB experiments \cite[Table~6-2]{Abazajian:2016yjj}.
        }
        \label{fig:squares}
\end{figure}

Figures~\ref{fig:derived_priors} and \ref{fig:joint_distro} were obtained using the slow roll approximation with a fixed number of $e$-folds, rather than  a self-consistent solution of the  matching equation.  For an extreme post-inflationary equation of state the pivot scale may be pushed to larger values of $N$, which typically reduces the running. However, Figure~\ref{fig:squares} shows the allowed regions of parameter space for two specified sets of bounds on the spectral parameters obtained from self-consistent solutions to the matching equation. These results are consistent with   Figure~\ref{fig:joint_distro}  -- if $\alpha \approx -0.002$ and $r<0.001$  a nontrivial region of  parameter space (if $1-\Delta$ is drawn from a logarithmic prior) would be consistent with these observations, but if $|\alpha| \approx 0.001$ the overall model space is excluded.

\afterpage{\clearpage}

In contrast to the small field case, the large field scenarios are likely to need no significant tuning of their initial conditions. Matching the overall amplitude of the perturbations will constrain $\lambda$ but inflation  occurs for generic values of $\Delta$ and $M$.

\section{Discussion}
\label{sec:discussion}

We have reviewed the inflationary scenarios generated by a 4-th order polynomial potential. This topic was first addressed in 1990  \cite{Hodges:1989dw,Hodges:1990bf} and has been the subject of numerous subsequent analyses.  Our treatment is based on a parametrization of $V(\phi)$ which makes it easy to identify the  inflationary regimes the potential supports and adds to previous discussions by including the spectral running among the possible observables. 

The  simplest  potentials, both of which are at odds with observations, are the quadratic and quartic models.  The 4-th order polynomial is the next-most-simple scenario that is bounded below. It also marks a critical threshold in the  inflationary model space as it is the most complex potential consistent with a renormalizable scalar field theory.  This potential has far richer phenomenology than the simplest models --  we catalogued eight different supported inflationary regimes (including the small field case). However, this scenario is not arbitrarily complex and many combinations of the spectral parameters $n_s$, $r$ and $\alpha_s$ cannot be generated by any configuration of the potential.

Interestingly,  the 4-th order model has previously been used to explore questions of tuning and naturalness in inflationary models \cite{Boyle:2005ug,Bird:2008cp,Boyle:2008ri,Ijjas:2013vea}. Our treatment allows us to give a quantitative and fully Bayesian assessment for the naturalness of any combinations of cosmological observables this potential can produce. As part of our analysis we have written down priors that describe the distributions of free parameters in the 4-th order scenario, looking at both the large field and small field regimes. The free parameters in the potential can be viewed as hyperparameters in a generative model for the spectral parameters and the resulting prior distributions are computed in Section~5.2, with examples  plotted in Figures~\ref{fig:derived_priors} and \ref{fig:joint_distro}.  

In the large field scenarios the joint distributions for $n_s$ and $r$ are sharply peaked  around the values expected from quadratic inflation but a wide range of tensor amplitudes can be generated within these scenarios. This is clearly true for the large field log prior example but even with a uniform prior case there is still non-zero support for small values of $r$; these cases are somewhat disfavoured relative to those with $r\gtrsim 0.1$ but the level of tuning involved to produce a tensor amplitude in the range $0.001 \lesssim r \lesssim 0.1$ would not be outlandish from this perspective.

Conversely, this analysis has shown that there are combinations of $n_s$, $r$ and $\alpha_s$ which cannot be produced by a 4-th order potential for any choice of parameter values.
In particular, given current limits on $n_s$, if $r$ and $\alpha_s$ are simultaneously constrained to have magnitudes smaller than $10^{-3}$  no scenarios we have identified within this potential would survive.
Such a result that would represent a significant threshold in the understanding of inflationary phenomenology, and thus provides a target for the designers of future experiments.

It is worth considering what we would learn if future experiments do  rule out the 4-th order potential. Clearly, if algebraic simplicity is held to be synonymous with naturalness this  would  diminish the credibility of the inflationary paradigm.  However,  the only small-field (i.e. $\phi \lesssim \mpl$ at all times) scenario that the 4th-order potential supports is the inflection point scenario described in Section~3, for which the parameters in the potential and the initial field configuration are both  highly tuned. Conversely,  potentials with a higher degree of algebraic complexity can support inflation at lower scales (and thus smaller field excursions) without fine-tuned initial conditions~\cite{East:2015ggf}. These potentials typically have large plateau, which can be a consequence of possible symmetries in the underlying theory of high energy particle physics but cannot be constructed from a quartic polynomial. If  natural models are those where the inflationary dynamics is the consequence of a fundamental symmetry rather than happenstance, the most natural small-field scenarios are {\em necessarily} more complex than the 4-th order polynomial, and metrics based on naturalness as opposed to simplicity will yield contradictory conclusions about the likelihood of inflation.    Conversely, the 4-th order potential supports large-field inflation without needing dramatic tunings to either the potential parameters or initial state. However, the robustness of this potential against corrections from Planck-scale operators is again a question of high energy physics rather than the simplicity of its algebraic form \cite{Lyth:1996im,Baumann:2014nda}. Consequently, the one inference we can draw is that in all cases the naturalness or prior likelihood of an inflationary scenario is best assessed in terms of the theory or theories of fundamental physics that are hypothesised to give rise to its potential, rather than the form of the potential itself. 

%

 \acknowledgments  We thank Shaun Hotchkiss, Kei-ichi Maeda, Mason Ng, and Hiranya Peiris for useful conversations on this topic. Part of this work was completed during a visit to the Centro de Ciencias de Benasque Pedro Pascual [RE]. 
 We made use of Matplotlib, NumPy and SymPy \cite{Hunter:2007ouj,vanderWalt:2011bqk,Meurer:2017yhf}.

\bibliographystyle{JHEP}
\bibliography{bib,autobib} 

\end{document}